\documentclass[twocolumn,aps,superscriptaddress,pre,10pt]{revtex4-2}

\usepackage[latin1]{inputenc}
\usepackage[T1]{fontenc}
\usepackage{setspace}
\usepackage{tabularx}
\usepackage{graphicx}
\usepackage{frontespizio}
\usepackage{amsmath}
\DeclareMathOperator{\sech}{sech}
\usepackage{amsthm}
\usepackage{amsfonts}
\usepackage{mathrsfs}
\usepackage{txfonts}
\usepackage{float}
\usepackage{braket}
\usepackage{booktabs}
\usepackage{xcolor}
\usepackage{natbib}
\usepackage{appendix}
\newenvironment{sistema}
  {\left\lbrace\begin{array}{@{}l@{}}}
  {\end{array}\right.}

\begin{document}

\title{A quantitative analysis of a generalized Hopfield model that\\
stores and retrieves mismatched memory patterns}

\author{Luca Leuzzi}
\affiliation{CNR-NANOTEC, Institute of Nanotechnology, Rome Unit, Piazzale Aldo Moro 5 - 00185 Roma, Italy}
\affiliation{University of Rome "La Sapienza",  Department of Physics, Piazzale Aldo Moro 5 - 00185 Roma, Italy}

\author{Alberto Patti}
\affiliation{University of Rome "La Sapienza",  Department of Physics, Piazzale Aldo Moro 5 - 00185 Roma, Italy}
\affiliation{IIT, Center for Life Nano- \& Neuro-Science,  Viale Regina Elena, 291 - 00161 Roma, Italy}
\email{alberto.patti@uniroma1.it}

\author{Federico Ricci-Tersenghi}
\affiliation{University of Rome "La Sapienza",  Department of Physics, Piazzale Aldo Moro 5 - 00185 Roma, Italy}
\affiliation{CNR-NANOTEC, Institute of Nanotechnology, Rome Unit, Piazzale Aldo Moro 5 - 00185 Roma, Italy}
\affiliation{INFN, sezione di Roma1, Piazzale Aldo Moro 5 - 00185 Roma, Italy}

\date{\today}

\begin{abstract}
We study a class of Hopfield models where the memories are represented by a mixture of Gaussian and binary variables and the neurons are Ising spins. We study the properties of this family of models as the relative weight of the two kinds of variables in the patterns varies. We quantitatively determine how the retrieval phase squeezes towards zero as the memory patterns contain a larger fraction of mismatched variables.
As the memory is purely Gaussian retrieval is lost for any positive storage capacity.  
It is shown that this comes about because of the spherical symmetry of the free energy in the Gaussian case.
Introducing two different memory pattern overlaps between spin configurations and each contribution to the pattern from the two kinds of variables one can observe that the Gaussian parts of the patterns act as a noise, making retrieval more difficult.
The basins of attraction of the states, the accuracy of the retrieval and the storage capacity are studied by means of Monte Carlo numerical simulations.
We uncover that even in the limit where the network capacity shrinks to zero, the (few) retrieval states maintain a large basin of attraction and large overlaps with the mismatched patterns. So the network can be used for retrieval, but with a very small capacity.
\end{abstract}

\maketitle

\section{Introduction}
\label{section:I}

The Hopfield model~\cite{Hopfield:article} is a model of associative memory and it has been used as a prototype of artificial neural networks that, because of the increased computational power, have recently been widely exploited as the most powerful and flexible tools of machine learning ~\cite{Ramsauer:article,Benedetti:article}. It is, then, of great importance to understand the mechanisms underlying models of neural networks. One of such mechanisms, that requires a deeper understanding, is the ability of a neural network model to learn \emph{mismatched} patterns, that is patterns made of variables of a different nature than those of the model. For examples, the classical Hopfield model is defined in terms of discrete variables (Ising spins), but the performance of the model when the memory to recover has a different kind of variables is not yet fully clear. This is the well known mismatching problem already studied in ~\cite{Bovier:article, Enter:article, Barra:article, Barra1:article} and that it is relevant in realistic applications ~\cite{Ruocco:article}.

Technically speaking, the Hopfield model describes the activity of a network of $N$ neurons when $P$ memories are stored in the neural network. The neurons are typified   by means of Ising spins, $s_i$, taking value $1$ if the neuron is active, $-1$ if the neuron is passive. The spins interact through the interaction matrix elements $J_{ij}$ that represent the synaptic efficacies between neurons $i$ and $j$. Their values are random, symmetric and quenched with respect to the dynamics of the neuronal activity but not independent. The Hamiltonian of such a system is given by:
\begin{equation}
H=-\frac{1}{2}\sum_{i\neq j}^{1,N} J_{ij}s_is_j\;.
\label{def:hamiltonian}
\end{equation}
In order to describe an associative memory, the $J_{ij}$ are built in terms of the $P$ stored memory patterns $\xi^{(\mu)}_i=\pm 1$ following the Hebb's learning rule \cite{Little:article,LittleShaw:article,Hebb:article}
\begin{equation}
J_{ij}=\frac{1}{N}\sum_{\mu=1}^{P} \xi_{i}^{(\mu)} \xi_{j}^{(\mu)}\;.
\label{def:J}
\end{equation}

The number $P$ of memorized patterns $\xi^{(\mu)}$ is such that $$\alpha=\lim_{N\rightarrow\infty}\frac{P}{N}$$ is the storage capacity: $\alpha = 0$ means that $P$ does not scale with the size $N$ of the neural network, whereas for $\alpha >0$ the amount of memories that can be retrieved from the neural network grows linearly with the number of neurons. 
The patterns are memorized in the sense that, in the noiseless situation, the optimal neuronal configurations, i.e.\ the minima of Eq.~(\ref{def:hamiltonian}), satisfy $s_i=\xi_i^{(\mu)}$ and the pattern $\xi^{(\mu)}$ is recovered.

Such a model has a phase diagram in the temperature ($T$) vs.\ $\alpha$ plane, displaying a high temperature paramagnetic phase (PM), a low temperature, high capacity spin glass phase (SG) and a low temperature, low capacity memory recovery phase (MR). The PM-SG transition is a second order one, whereas the transition between SG and MR is a first order one, with a spinodal line signaling the appearance of the retrieval states, i.e., the states of the memorized patterns.
In the MR phase the model is able to serve as an associative memory model. 

This picture changes if we consider the memories as made of continuous, Gaussian distributed, variables
\begin{equation}
p(\xi)= \frac{1}{\sqrt{2\pi}}e^{-\frac{\xi^2}{2}}\;.
\label{eq:gaussiandist}
\end{equation}
In this case, first studied in Ref.~\cite{Amit2:article}, the model presents retrieval only when $\alpha=0$, i.e.\ if the number of memorized patterns grows less than linearly in the thermodynamic limit. On the contrary, if the number of memories grows like $N$, retrieval is lost. This lack of retrieval might be explained by the spherical symmetry in the free energy that the continuous memory variables introduce. As soon as the number of patterns is allowed to be extensive, the system does not know how to distinguish the Gaussian memories. 

In order to achieve a deeper understanding of the situation where patterns and variables are partially mismatched we follow the work of ~\cite{Barra:article,Barra1:article} by studying a family of mixed Hopfield models (MHM), where each memory has some bimodal ($\pm 1$) variables and some Gaussian variables, and we study what happens to retrieval as the relative number of the different kinds of variables in a memorized pattern changes. 
Calling $p$ the fraction of continuous memory variables, each pattern $\xi^{(\mu)}$ has $(1-p) N$ variables drawn from a bimodal distribution and $pN$ real continuous variables, drawn from distribution (\ref{eq:gaussiandist}). This is not the only possible choice to study mismatch in the Hopfield model: in~\cite{Agliari:article} a sub-extensive set of bimodal memories and an extensive set of Gaussian ones have been considered, obtaining a phase diagram similar to the one of the standard Hopfield model.

We summarize the main results reported in this work, while describing the organization of the manuscript. In Sec.~\ref{section:II} we report the free energy of the model and the corresponding saddle point equations. Although the phase diagram was already known~\cite{Barra:article,Barra1:article}, we focus on the quantitative dependence of critical lines on the fraction $p$ of mismatched patterns.
In particular at zero temperature we find critical lines to shrink as $(1-p)^2$.
We also discuss in great detail the solution in the $\alpha=0$ limit, since that limit is the only relevant one for a large fraction of mismatched patterns.
In Sec.~\ref{subsection:IIF} we work out a replica symmetric theory for the mixed model with two different order parameters, one for the Gaussian variables and one for bimodal variables. While the bimodal variables feel a local field consisting of a ferromagnetic-like signal and a spin-glass noise, the Gaussian variables only feel a spin-glass noise generated by both the non-retrieved patterns (as in the original Hopfield model) {\em and} the continuous contribution of the pattern to be retrieved. Notwithstanding this, we obtain that both the Gaussian and the bimodal overlaps remain very large until the solution is lost at the spinodal point.

In Sec.~\ref{section:III} we perform a numerical analysis of the mixed Hopfield model using a Metropolis algorithm at zero temperature. We study the basin of attraction of the retrieval states and we calculate both the minimal overlap required to achieve retrieval and the accuracy in retrieval. The minimal initial overlap mostly depends on $\alpha$, while the retrieval accuracy mostly depends on $p$. Also the numerical results support the idea that, even for a large fraction of mismatched pattern variables, the retrieval state have a large basin of attraction and large retrieval accuracy. So the model can be used for storage and retrieval also in that limit.
At last, following the work of ~\cite{Amit:article} and ~\cite{Stiefvater:article}, we numerically compute $\alpha_c(p)$ and verify that it is larger than the analytical prediction based on the replica symmetric solution for any $p$ value, while keeping the $(1-p)^2$ dependence on the fraction of mismatched variables.

\section{The mixed Hopfield model at equilibrium}
\label{section:II}

\subsection{Free energy of the mixed Hopfield model}
\label{subsection:IIA}
We want to study the fully connected mixed Hopfield model, i.e., a model of $N$ bimodal neurons storing $P$ patterns in memory. Each learnt pattern $\xi^{(\mu)}$ is made up of $pN$, $p<1$, random, independent Gaussian variables and $(1-p)N$ random, independent bimodal variables. The Hamiltonian of such a system can be written as Eqs. (\ref{def:hamiltonian},\ref{def:J}).

Since we are interested in the properties of the model near saturation, we allow the number $P$ of patterns to
diverge in the thermodynamic limit: their number is $P=\alpha N$ where
$\alpha$ is a finite number. We also suppose the existence of a finite
number $s$ of
aligned patterns $\xi^{(\nu)}$, with $\nu=1, \ldots, s$,  and we add the contribution of conjugate fields to these patterns in the Hamiltonian:
\begin{equation}
H_h=-\sum_{\nu=1}^{s}h^{(\nu)}\sum_i\xi_{i}^{(\nu)}s_i
\end{equation}
In order to calculate the average free energy per spin, $f$, we use the  replica method ~\cite{MPV:book}:
\begin{equation}
f=\lim_{n\rightarrow 0}\lim_{N\rightarrow \infty}-\frac{1}{\beta n N}(\mathbb{E}[Z^n]-1)
\label{eq:freenergy}
\end{equation} 
where $n$ is the number of replicas, $\mathbb{E}$ is the average over the random patterns and $Z$ is the partition function. The details of the computation are reported in Appendix \ref{app:A}. Here, we give the result of the free energy per spin in the replica symmetric theory at inverse temperature $\beta$
\begin{eqnarray}
\label{eq:FE1}
&&\beta f=   \frac{\beta \alpha}{2} + \frac{\beta}{2}\sum_{\nu}\left(m^{(\nu)}\right)^2+\frac{\alpha}{2}\ln(1-\beta+\beta q)\\
&&\quad -\frac{\alpha}{2}\frac{\beta q}{1-\beta+\beta q} +\frac{\alpha\beta^2r}{2}(1-q)
\nonumber 
\\
\nonumber
&&\quad -\int {\mathcal{D}} z\left\{p
\ln\left(2\cosh\left(z\beta\sqrt{\alpha r+\sum_\nu (m^{(\nu)}+h^{(\nu)})^2}\right)\right) \right.\\
\nonumber
&&\quad  +(1-p) \left.
\mathbb{E}_B\left[ \ln2\cosh\left( \beta z \sqrt{\alpha r}+\beta \sum_\nu (m^{(\nu)}+h^{(\nu)}) \xi_{B}^{(\nu)}\right)\right]\right\}
\nonumber
\\
\nonumber
&& \mbox{ with    } {\mathcal{D}} z\equiv \frac{dz}{\sqrt{2\pi}}e^{-\frac{z^2}{2}}.
\end{eqnarray}

The free energy is written in terms of the order parameters of the theory:  the overlap $m^{(\nu)}$ of a configuration with one aligned memory $\nu$,  the overlap $q$ between two configurations belonging to two independent replicas and the spin glass noise $r$, that is a function of $q$ and it is caused by the presence of an infinite number of non aligned patterns.
Considering, without loss of generality,  one aligned pattern $\xi^{(1)}=\xi$, with $m^{(1)}=m$, and sending the external field to zero, we can obtain the saddle point equations for this model

\begin{eqnarray}
m&=&p \, m\, \beta\int {\mathcal D}z\,  \sech^2\left(z\beta\sqrt{\alpha r+m^2}\right) 
\label{eqn:SDEm}
\\
\nonumber
&&+(1-p)\int {\mathcal{D}}z 
\, \mathbb{E}_B\Big[\xi\tanh\beta\left(z\sqrt{\alpha r}+m\xi\right)\Big].
\end{eqnarray}
\begin{eqnarray}
\label{eqn:SDEq}
q&=&p\int {\mathcal D}z\, \tanh^2\left(z\beta\sqrt{\alpha r+m^2}\right) +\\
\nonumber
&&+(1-p)\int  {\mathcal D}z\,  \mathbb{E}_B\left[\tanh^2\left(z\beta\sqrt{\alpha r}+\beta \xi m\right)\right]
\end{eqnarray}

\begin{eqnarray}
r&=&\frac{q}{\left[]1-\beta(1-q)\right]^2}
\label{eqn:SDEr}
\end{eqnarray}
We stress that in the case of patterns with  purely Gaussian entries,  $p=1$, assuming $m\neq0$ in Eq.~(\ref{eqn:SDEm}), yields  $1=\beta(1-q)$ which causes Eq.~(\ref{eqn:SDEr}) to be ill defined. A well defined theory is recovered only if we had considered $\alpha=0$ in the beginning, that is, $P$ scaling less than $O(N)$. Indeed, in this case there is no need to introduce the variable $r$ which, in fact, represents the random overlaps with the misaligned patterns, acting as an overall Gaussian noise. A description of the Gaussian Hopfield model at $\alpha=0$ can be found in ~\cite{Amit2:article}.

\subsection{The mixed model near saturation at finite temperature}
\label{subsection:IIE}
The phase diagram of the mixed model is very similar to the standard one, but with a few differences. Just like in the standard model \cite{Amit:article} we have:
\begin{itemize}
\item the paramagnetic (PM) phase, in which $m=q=0$. In this phase, the noise given by the temperature is too high for the neurons to have any collective behavior;
\item the pure spin glass (SG)  phase, with $q\neq 0$ and $m=0$. Here the combined effect of temperature and high number of patterns does not allow the system to serve as an associative memory with retrieval;
\item the metastable retrieval phase. Here a spin glass phase and a ferromagnetic-like ``memory recovery'' (MR) phase ($m\neq0$ but $q=0$) coexist, but the retrieval states are metastable and the appearance of such states is signaled by the spinodal lines reported in \ref{subsubsection:2};
\item the pure retrieval phase. Undergoing a first order phase transition the MR phase becomes the stable phase. This happens in the part of the phase diagram delimited by the lines studied in section \ref{subsubsection:3}.
\end{itemize}

\subsubsection{Second order PM-SG transition}

The transition between the paramagnetic and the spin glass phase is of the second order and the transition temperature can, thus, be computed using Eqs. (\ref{eqn:SDEq})-(\ref{eqn:SDEr}) with  $m=0$ and expanding in powers of $r$ and $q$:
\begin{eqnarray}
r&\simeq& \frac{q}{(1-\beta)^2},  \\
q&\simeq& \beta^2 \alpha r .
\end{eqnarray}
The transition temperature
\[
T_g=1+\sqrt{\alpha} 
\]
is the one below which a $q\neq 0$ solution continuously arises. 
This is the same as in the standard Hopfield model ~\cite{Amit:article} since the saddle point equations with $m=0$ do not depend on the fraction  $p$ of Gaussian variables in a memory pattern. This result coincides with the results found in \cite{Barra:article} in the context of restricted Boltzmann machines  \cite{RBM1:article, RBM2:article}.

\subsubsection{Spinodal curves}
\label{subsubsection:2}
Next we want to draw the spinodal line that signals the appearance of the retrieval state as a metastable state. In fact, at the spinodal points, a new minimum of the free energy at $m\neq0$ appears. Such minimum has higher free energy than the SG one, but as temperature $T$ or storage $\alpha$ are lowered the free energy difference becomes smaller and the SG minimum eventually becomes the metastable one and a first order phase transition to MR occurs. Spinodals are still rather important because, although at this stage MR is not thermodynamically dominant, a system starting off its dynamics with a configuration similar enough to the one of a stored memory will fall into the retrieval minimum and will never leave it in the $N\rightarrow\infty$ limit in a mean field theory, even in the presence of thermal noise. This is also the case of a finite size system at zero temperature as will be studied in Section \ref{section:III}. In order to determine the spinodal curves we, first, simplify the notation  introducing the following definitions:
 \begin{eqnarray}
I_k(m,q)&\equiv&\int \mathcal{D}z\ \tanh^k\left(\beta m +\beta z\sqrt{\alpha\, r} \right)
\\
J_k(m, q)&\equiv&\int \mathcal{D}z\ \tanh^k\left(\beta z\sqrt{m^2+\alpha r }\right),
\end{eqnarray} 
in such a way that Eqs. (\ref{eqn:SDEm})-(\ref{eqn:SDEq}) can be rewritten as:
 \begin{eqnarray}
m&=&(1-p)I_1(m,q)
+p\, m\,\beta\left(1-J_2(m,q) \right) 
\\
q&=&(1-p)I_2(m,q)
+p\, J_2(m,q).
\end{eqnarray} 

\begin{figure}[t!]
\centering
\includegraphics[width=0.99\columnwidth]{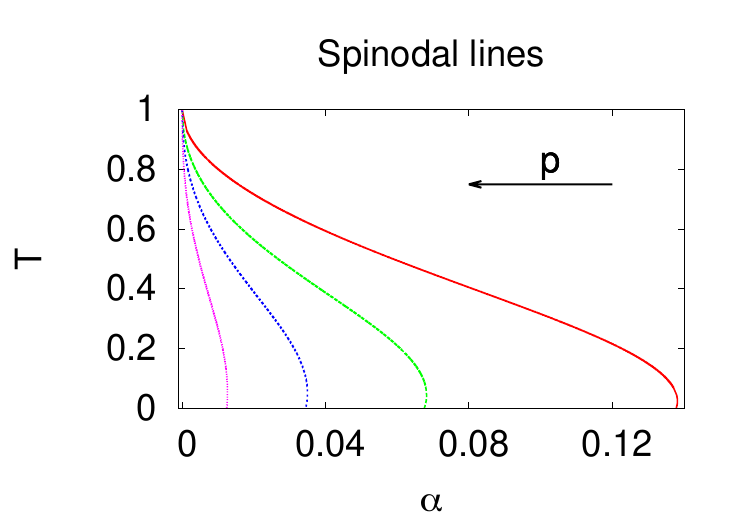}
\caption{Spinodal lines of the mixed model for various values of $p$. From right to left, the  values of $p$ are $p=0$ (red), $p=0.3$ (green), $p=0.5$ (blue), $p=0.7$ (purple).}
\label{fig:spinodali}
\end{figure}

In the following we will omit to write the arguments of the integrals $I_k$ and $J_k$, because they will always be the same as in the previous equations. 
With the procedure reported in Appendix \ref{app:spineT} we numerically compute the retrieval spinodal lines $\alpha_c(p,T)$ plotted in Fig.~\ref{fig:spinodali}. This method allows us to find quantitative information on how the spinodal lines $T(\alpha)$ change as functions of $(1-p)$.

\begin{figure}[t!]
\includegraphics[width=0.88\columnwidth]{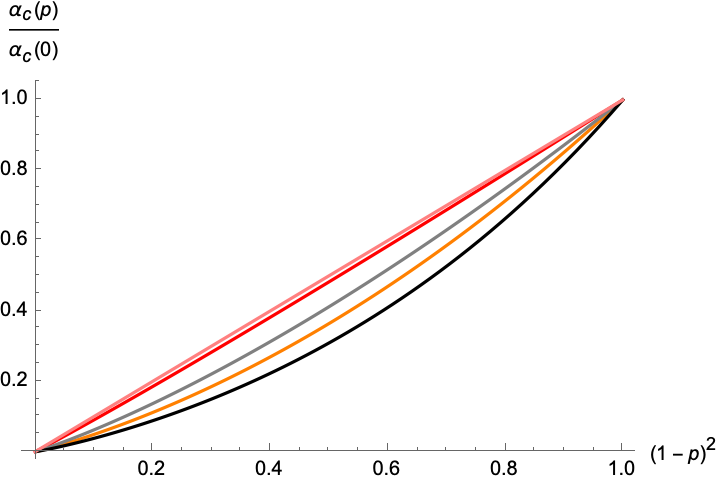} 
\caption{Behavior of the saturation points $\alpha_c(p)/\alpha_c(0)$ as a function  of $(1-p)^2$ for different values of the temperature: from bottom to top, $T=0.955$ (black), $T=0.832$ (orange), $T=0.7$ (gray), $T=0.414$ (red) and $T=0$ (pink). The latter is a straight line.}
\label{fig:alphaspinodale}
\end{figure}

In Fig.~\ref{fig:alphaspinodale} we plot $\alpha_c(p,T)/\alpha_c(0,T)$ as a function of $(1-p)^2$. It is clear how, as temperature approaches zero, the curves tend to linearize getting closer to the bisector.  This will be found as an exact result in Sec. \ref{subsection:IID} where we will consider  the zero temperature limit. 

From these curves we can draw the spinodal lines by means of a fit with linear and quadratic terms. The result of this procedure, outlined in Appendix \ref{app:C1}, is depicted in Fig.~\ref{fig:spinodali}.
It is clear that the more $p$ grows, the greater is the portion of the phase diagram where  retrieval is not allowed, in agreement with \cite{Barra:article, Barra1:article} where the phase diagram of restricted Boltzmann machines with generic priors is found. 
Also, as $p$ approaches $1$,  the spinodal lines get closer and closer to the vertical line $\alpha=0$.
This accounts for the fact that in the purely Gaussian case, retrieval states do not exist as soon as the number of patterns scales with the number of neurons of the network.

\subsubsection{First order transition curves}
\label{subsubsection:3}
As temperature decreases a first order phase transition occurs in which the  metastable retrieval states become thermodynamically dominant. In order to find the transition line on the ($T$,$\alpha$) phase diagram, we follow  a method similar to the one used for the computation of the spinodal, that we report in Appendix \ref{app:spineT}, where we equal the free energies in the two phases.

\begin{figure}[t!]
\centering
\includegraphics[width=0.99\columnwidth]{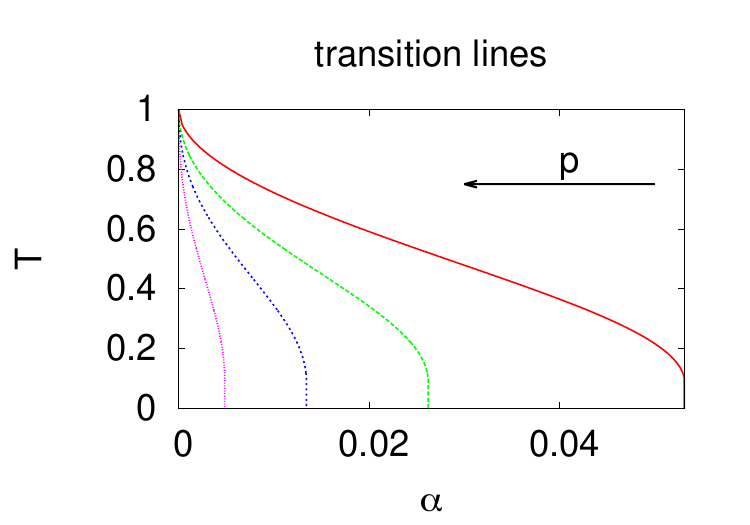}
\caption{First order transition lines of the mixed model for various values of $p$. From left to right, we use $p=0$ (red), $p=0.3$ (green), $p=0.5$ (blue), $p=0.7$ (purple).}
\label{fig:transizione}
\end{figure}

\begin{figure}[t]
\includegraphics[width=0.88\columnwidth]{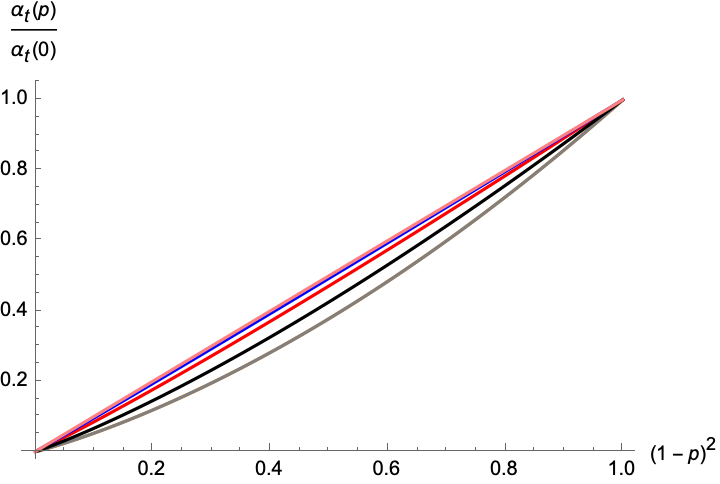}
\caption{The behavior of $\alpha_t(p)/\alpha_t(0)$ as a function of $(1-p)^2$ is very smooth and tends to a straight line in the limit $T=0$ (upper line).}
\label{fig:alphat}
\end{figure}

The results are shown in Figs.~\ref{fig:transizione} and \ref{fig:alphat}.
In the latter we plot $\alpha_t(p,0)/\alpha_t(0,0)$ as a function  of $(1-p)^2$ for various values of the temperature. Here we call $\alpha_t(p, T)$ the critical value of $\alpha$ at the transition for each value of $p$ and $T$.

The behavior in $p$ is very similar to the case of the spinodals: $\alpha_t(p,0)/\alpha_t(0,0)$ tends to behave exactly as $(1-p)^2$  as the temperature reaches zero. As we report in Appendix \ref{app:C1} we can draw the transition lines that are showed in Fig.~\ref{fig:transizione}. It is worth noticing that, again, the lines in the phase diagram are squeezed in the limit $\alpha=0$ as we approach the purely Gaussian model and that the described method allows us to  determine how the first order recovery lines change as functions of $(1-p)$.

\subsection{Mixed model at $\alpha=0$ and the Mattis states}
\label{subsection:IIC}

The case $\alpha=0$ is the one in which there is only a finite number of memorized patterns $P$. Here we take into account  the MR phase made of so-called Mattis states ~\cite{Mattis:article}, i.e., the states having non vanishing overlap with only one of the patterns. At $\alpha=0$ there is only one independent saddle point equation (because $q=m^2$):
\begin{equation}
\begin{split}
&m=p\, m\, \beta \int {\mathcal{D}} z\,  \sech^2\left(z\beta m\right)
+(1-p)\tanh(\beta m) .
\label{eqn:SPDm0}
\end{split}
\end{equation}

As the transition from the paramagnetic phase to the retrieval phase at $\alpha=0$ is a second order one, we can expand Eq. (\ref{eqn:SPDm0}) for small values of $m$ obtaining:
\begin{equation}
m=\beta m -\frac{1}{3}m^3\beta^3(1+2p) +O(m^4).
\end{equation}
For $\beta<1$ there is only the paramagnetic solution, $m=0$, while for $\beta>1$ the Mattis states appear and the transition temperature $T=1$ is independent on the value of $p$. 

\begin{figure}[t!]
\centering
\includegraphics[width=0.99\columnwidth]{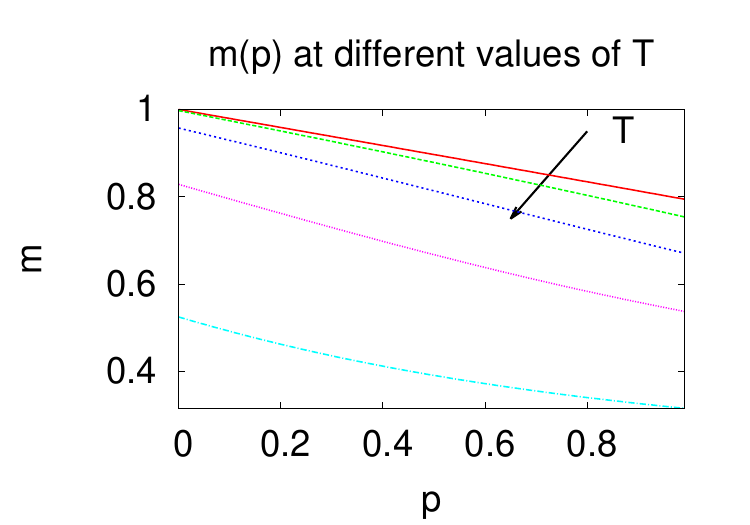}
\caption{Plot of the overlap $m$ as a function of $p$ at different values of the temperature for $\alpha=0$. Following the arrow, the used values of $T$ are $T=0.1$ (red), $T=0.3$ (green), $T=0.5$ (blue), $T=0.7$ (purple),  $T=0.9$ (cyan).}
\label{fig:mvsp}
\end{figure}

\begin{figure}[t!]
\centering
\includegraphics[width=0.99\columnwidth]{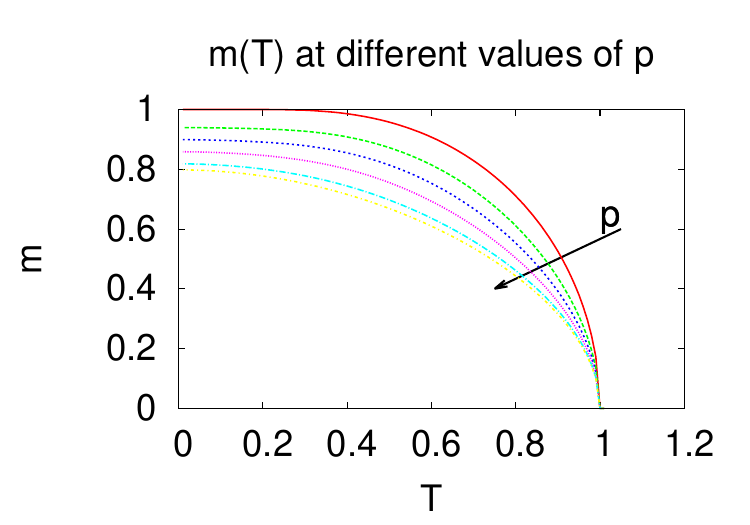} 
\caption{Plot of the memory overlap $m$ as a function of the temperature at different values of $p$ for $\alpha=0$. Following the arrow, the used values of $p$ are $p=0$ (red), $p=0.3$ (green), $p=0.5$ (blue), $p=0.7$ (purple),  $p=0.9$ (cyan),  $p=1$ (yellow).}
\label{fig:mvsT}
\end{figure}

It is useful to recover the expressions at zero temperature, using the following limit:
\begin{equation}
\lim_{\beta\rightarrow\infty}\beta\left(1-\tanh^2(\beta\, x)\right)=2\, \delta(x)  
\label{eqn:limite}
\end{equation}
By calling $m_0=\lim_{\beta\rightarrow \infty} m$, we can take the zero temperature limit of Eq.~(\ref{eqn:SPDm0}), obtaining
\begin{equation}
m_0=\pm \left[1-p\left(1-\sqrt{\frac{2}{\pi}}\right)\right].
\end{equation}
In particular, in the purely Gaussian pattern model \cite{Amit2:article}, it is
$$|m_0|=\sqrt{\frac{2}{\pi}}.$$ As it could be expected, the overlap diminishes as $p$ grows.  At zero temperature the decrease is linear. 
For small values of $T$ a self-consistency expression of $m$ can be computed as
\begin{multline}
m=1-2e^{-2m\beta}\\
+p\left[-1+2e^{-2m\beta}- \sqrt{\frac{2}{\pi}}+4m\beta e^{2(m\beta)^2}\mbox{Erfc}\left(\sqrt{2}m\beta\right)\right]
\end{multline}
where
\begin{equation}
\mbox{Erfc}(x)=1-\mbox{Erf}(x)=\frac{2}{\sqrt{\pi}}\int_{x}^{\infty} dz\,  e^{-z^2}.
\end{equation}
The solution $m(p,T)$ to the above equations is displayed in Figs.~\ref{fig:mvsp} and  \ref{fig:mvsT}. In Fig.~\ref{fig:mvsp} we look at the overlap $m$ as a function of $p$ for different values of the temperature. It is clear that the zero temperature limit is recovered (linear behavior) and that $m$ gets smaller and smaller as $p$ grows, but the behavior slowly loses its linearity as temperature approaches $T=1$. The behavior of  $m$ as a function of $T$ can be seen in Fig.~\ref{fig:mvsT}, where it can be noticed that as $p$ increases, the values of $m$ at each temperature decrease, i.e., the overlap with the memory gets smaller as soon as we add more and more Gaussian spins to the pattern, but we can still have the retrieval phase for every value of $p$. We also see that the dependence on $p$ gets stronger at smaller temperatures, while it gets weaker as we approach the critical point. 



\subsection{The mixed model near saturation at $T=0$}
\label{subsection:IID}
At this point it can be of interest to find what is the {\em storage capacity} of the network, i.e., how many patterns can be memorized in order to maintain retrieval possible. To this end we allow $\alpha > 0$ and study the general saddle point equations (\ref{eqn:SDEm}), (\ref{eqn:SDEq}) and (\ref{eqn:SDEr}).

Here, we take into account the zero temperature case. Therefore we rewrite (\ref{eqn:SDEm}) and (\ref{eqn:SDEq}) for $T=0$. In the case of  (\ref{eqn:SDEq}) it is useful to define the local susceptibility $\chi\equiv \beta(1-q)$. Calling $\lim_{\beta\rightarrow \infty}m=m_0$ and $\lim_{\beta\rightarrow \infty}\chi=\chi_0$, we  take the zero temperature limit using formula (\ref{eqn:limite}):
\begin{eqnarray}
\chi_0&=&\sqrt{\frac{2}{\pi}}\frac{1-\chi_0}{\sqrt{\alpha}}\left[\frac{p}{\sqrt{1+x^2}} + (1-p)\, e^{-\frac{x^2}{2}}
\right], 
\label{eqn:chi0}
\end{eqnarray}
where \begin{eqnarray}
x&\equiv&\frac{m_0(1-\chi_0)}{\sqrt{\alpha}}.
\label{def:x}
\end{eqnarray}
We also obtain the zero temperature limit of Eq.~(\ref{eqn:SDEm}):
\begin{equation}
m_0=\sqrt{\frac{2}{\pi}}
\, \frac{p\, x}{\sqrt{1+x^2}} 
+(1-p)\, \mbox{Erf}\left(\frac{x}{\sqrt{2}}\right)
\label{eqn:m0}
\end{equation}

The first thing we would like to know is how the storage capacity depends on the fraction $p$ of Gaussian variables in the  memory patterns. 
From Eqs. (\ref{eqn:chi0}) and  (\ref{eqn:m0}) we have 
\begin{equation}
\frac{m_0(1-\chi_0)}{1-p}= \frac{\sqrt{\alpha}\, x}{1-p}
=\mbox{Erf}\left(\frac{x}{\sqrt{2}}\right)-\sqrt{\frac{2}{\pi}}x\, e^{-\frac{x^2}{2}}
\end{equation}
The above equation has always the $x=0$ solution, that is $m_0=0$, absence of retrieval.
As  $\sqrt{\alpha}/(1-p)\leq  0.371356$, though, also solutions with $x\neq 0$ arise. Therefore we have the expression for the storage capacity as a function of the fraction $p$ of Gaussian variables:
\begin{equation}
\alpha_c(p)=\alpha_c(0)\, (1-p)^2, 
\label{eq:spine}
\end{equation}
with $\alpha_c(0) = 0.1379$ ~\cite{Amit:article}.
This equation tells us that as $p\rightarrow 1$ we lose quadratically the storage capacity. On the contrary, at $p=0$ the value of the original Hopfield model is found.

Eq. (\ref{eq:spine}) corresponds to the zero temperature spinodal line of the retrieval phase in the phase diagram $(p,\alpha)$ as can be verified 
defining the functions 
\begin{eqnarray}
w_1(m_0,\chi_0)&\equiv &m_0
-\sqrt{\frac{2}{\pi}}
\, \frac{p\, x}{\sqrt{1+x^2}} 
-(1-p)\, \mbox{Erf}\left(\frac{x}{\sqrt{2}}\right)
\nonumber
\\
w_2(m_0,\chi_0)&\equiv& \chi_0-
\sqrt{\frac{2}{\pi}}
\frac{x}{m_0}\left[\frac{p}{\sqrt{1+x^2}} + (1-p)\, e^{-\frac{x^2}{2}}
\right]
\nonumber
\end{eqnarray}
and numerically solving the system of equations
\begin{equation}
\begin{sistema}
w_1(m_0,\chi_0)=0\\
w_2(m_0,\chi_0)=0\\
\frac{\partial w_1}{\partial m_0}\frac{\partial w_2}{\partial \chi_0}-\frac{\partial w_1}{\partial \chi_0}\frac{\partial w_2}{\partial m_0}=0.
\end{sistema}
\label{eq:spine_sistem}
\end{equation}
The result of the numerical solution is the bisecting line in Fig.~\ref{fig:alphaspinodale}.

\subsection{Gaussian and bimodal contributions to the memory overlap}
\label{subsection:IIF}
We may ask what is the contribution to the total overlap, $m^{(\nu)}$, with the pattern $\nu$, of the Gaussian and of the bimodal variables separately. In fact, one may wonder if one of the two kinds of variables that form the patterns dominates retrieval, if both contribute in the best possible way they can or if the presence of one helps the other for the sake of memory retrieval. 
This task can be achieved by working out the partition function using different order parameters for the Gaussian and the bimodal parts. The calculations can be found in Appendix \ref{app:C}. Namely, we make use of the following definitions for the memory overlaps of the neuron configurations of replica $\rho$, $s^{(\rho)}$, with the bimodal -- $m^{(\nu)}_{\rho,B}$ --  and Gaussian -- $m^{(\nu)}_{\rho,G}$ -- contributions to the memory pattern $\xi^{(\nu)}$ and for the replica overlaps $q^{B}_{\rho\sigma}$, $q^{G}_{\rho\sigma}$ of the neurons of replicas $\rho$ and $\sigma$ coupled, respectively, to bimodal or Gaussian entries in the memories:

\begin{equation}
m_{\rho,G}^{(\nu)}=\frac{1}{pN}\sum_{i=0}^{pN-1}\xi_{i,G}^{(\nu)} s_{i}^{(\rho)}   
\end{equation}

\begin{equation}
m_{\rho,B}^{(\nu)}=\frac{1}{(1-p)N}\sum_{i=pN}^{N}\xi_{i,B}^{(\nu)} s_{i}^{(\rho)}    
\end{equation}

\begin{equation}
 q_{\rho\sigma}^G=\frac{1}{pN}\sum_{i=0}^{pN-1} s_{i}^{(\rho)} s_{i}^{(\sigma)} 
\end{equation}

\begin{equation}
q_{\rho\sigma}^B=\frac{1}{(1-p)N}\sum_{i=pN}^{N} s_{i}^{(\rho)} s_{i}^{(\sigma)}   
\end{equation}

And, by writing the corresponding Dirac's deltas in their Laplace transform, we call $r_{\rho\sigma}^G$ and $r_{\rho\sigma}^B$ the Lagrangian multipliers of respectively $q_{\rho\sigma}^G$ and $q_{\rho\sigma}^B$. Here, we report the free energy in the replica symmetric theory

\begin{equation}
\begin{split}
&f_{RS}=\frac{\alpha}{2}+ \frac{\alpha}{2\beta}\Bigg[\ln\left(1-\beta+\beta(pq^G+(1-p)q^B)\right)\\
&-\frac{\beta(pq^G+(1-p)q^B)}{1-\beta+\beta(pq^G+(1-p)q^B)}  \Bigg] \\
&-\frac{1}{2}\sum_{\nu}\left(pm_{G}^{(\nu)} +(1-p)m_{B}^{(\nu)} \right)^2-\frac{p\alpha \beta }{2}r^Gq^G\\
&-\frac{(1-p)\alpha \beta }{2}r^Bq^B +\frac{p}{2\beta }\sum_{\nu} \frac{\left(m_G^{(\nu)}\right)^2}{1-q^G}+(1-p)\sum_{\nu}m_{B}^{(\nu)} \lambda_{B}^{(\nu)} \\
&+\frac{\alpha\beta p}{2}r^G-\frac{p}{\beta}\int {\mathcal {D}}z\ln2\cosh\left(z\beta\sqrt{\alpha r^G}\right)+\\
&+\frac{(1-p)\alpha\beta r^B}{2}-\frac{1-p}{\beta}\int {\mathcal{D}}z \, 
\mathbb{E}_B\left[\ln2\cosh\left(\beta\tilde h_B\right)\right],
\label{eq:freeE}
\end{split}
\end{equation}
where we have defined the local field for the bimodal contribution to the memory patterns
\begin{eqnarray}
\tilde h_B \equiv z \sqrt{\alpha r^B}+\sum_\nu\lambda_{B}^{(\nu)}\xi_{B}^{(\nu)}.
\label{eq:h_B}
\end{eqnarray}
Next we can compute the saddle point equations in order to understand the thermodynamic behavior of the order parameters of such a theory. \\
The stationary states of the theory will then be given by the following equations:

\begin{equation}
q^B=\int {\mathcal{D}}z\, \mathbb{E}_B\left[\tanh^2\left(\beta\tilde h_B\right)\right]
\label{eq:qb}
\end{equation}

\begin{equation}
q^G=\int {\mathcal{D}}z\, \tanh^2\left(z\beta\sqrt{\alpha r^G}\right)
\label{eq:qg}
\end{equation}

\begin{equation}
m_B^{(\nu)}=\int{\mathcal{D}}z\, \mathbb{E}_B\left[\xi_{B}^\nu\tanh\left(\beta \tilde h_B\right)\right]
\label{eq:mb}
\end{equation}

\begin{equation}
 \lambda_B^{(\nu)}=pm_G^{(\nu)}+(1-p)m_B^{(\nu)}
\label{eq:lb}
\end{equation}

\begin{equation}
\begin{split}
m_G^{(\nu)}= \frac{\chi_G(1-p)m_B^{(\nu)}}{1-p\chi_G}
\label{eq:mg}
\end{split}
\end{equation}

\begin{equation}
\begin{split}
r^B=\frac{pq^G+(1-p)q^B}{\left[1-\beta+\beta(pq^G+(1-p)q^B)\right]^2}
\label{eq:rb}
\end{split}
\end{equation}

\begin{equation}
\begin{split}
r^G=r^B+\frac{1}{\alpha}\sum_{\nu=1}^s\frac{\left(m_G^{(\nu)}\right)^2}{\chi_G^2}
\label{eq:rg}
\end{split}
\end{equation}
\noindent where we have defined the local Gaussian susceptibility $$\chi_G=\beta(1-q^G).$$ 
Giving a first glance to these equations we can understand a few interesting features. For example, the local field $\tilde h_B$ in Eqs. (\ref{eq:qb}) and (\ref{eq:mb}) felt by the bimodal variables, is composed of two parts: 
a ``ferromagnetic'' part $$\sum_{\nu=1}^s\lambda_{B}^{(\nu)} \xi_{B}^{(\nu)}$$ and a spin glass part $$z\sqrt{\alpha r^B},$$ that is generated by the random overlaps with the misaligned patterns. 
Both the bimodal and Gaussian variables contribute to both parts of this local field, as stated by equations (\ref{eq:lb}) and  (\ref{eq:rb}). 
For what concerns the Gaussian contribution to the local field, from Eq. (\ref{eq:qg}) it can be noticed that the  ferromagnetic  contribution is absent. The Gaussian variables  only feel a Gaussian noise generated by both the unretrieved patterns and by the Gaussian contribution to the retrieved patterns. Such random normally distributed noise is always greater than the one we would have in the original Hopfield model, as can be seen in equation  (\ref{eq:rg}) whose second term in the right hand side of the equation is always non negative.
At last, equation  (\ref{eq:mg}) shows that the retrieval phase cannot exist in the purely Gaussian Hopfield model because as $p=1$ it is  $m_G=0$.

\begin{figure}[t!]
\centering
\includegraphics[width=0.99\columnwidth]{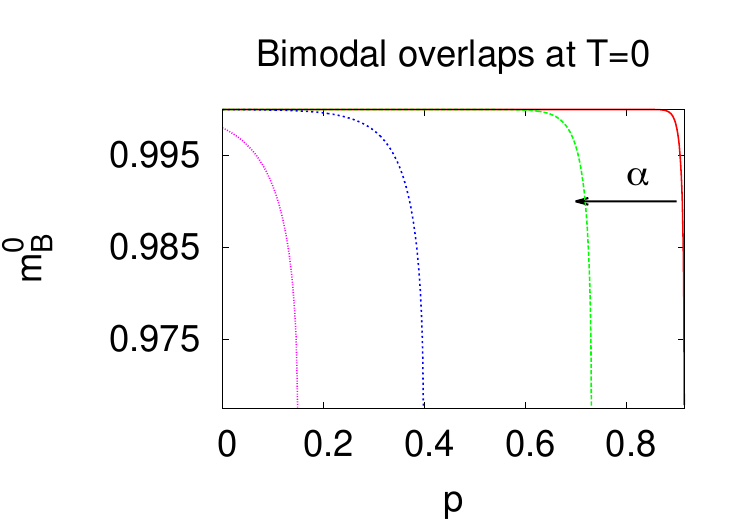}
\caption{The bimodal contribution to the memory overlap at zero temperature $m_B^0$ as a function of $p$ for different values of $\alpha$. Following the arrow, the used values of $\alpha$ are $\alpha=0.001$ (red), $\alpha=0.01$ (green), $\alpha=0.05$ (blue), $\alpha=0.1$ (purple).}
\label{fig:mb}
\includegraphics[width=0.99\columnwidth]{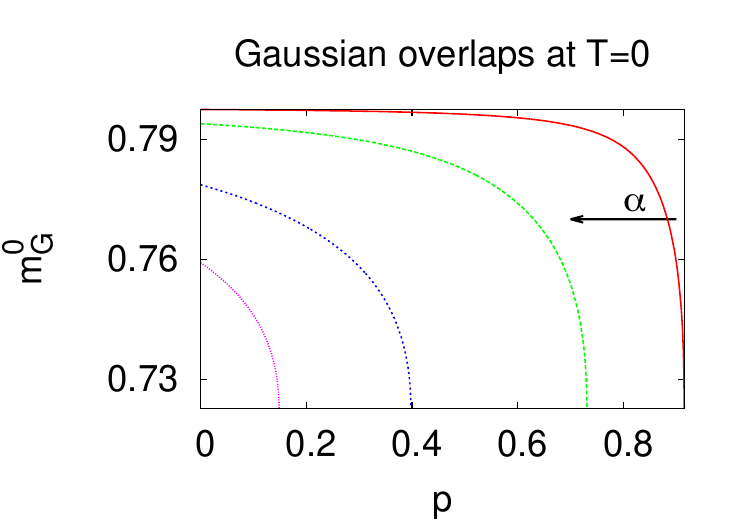} 
\caption{The Gaussian variables contribution to the memory overlap at zero temperature, $m_G^0$, as a function of $p$ for different values of $\alpha$. Following the arrow, the used values of $\alpha$ are $\alpha=0.001$ (red), $\alpha=0.01$ (green), $\alpha=0.05$ (blue), $\alpha=0.1$ (purple).}
\label{fig:mg} 
\end{figure}

\subsubsection{The model at T=0}

For a first analysis of the saddle point equations, it can be useful to look at the zero temperature limit. We will focus our analysis on the case where just one pattern is aligned ($s=1$).\\
We define the local bimodal susceptibility $$\chi_{B}=\beta(1-q^B),$$  the zero temperature limits of the local susceptibilities $$
\chi_{B}^0=\lim_{\beta\rightarrow \infty}\chi_{B},
\quad \chi_{G}^0=\lim_{\beta\rightarrow \infty}\chi_{G}$$ and of the memory overlaps $$m_{B}^0=\lim_{\beta\rightarrow \infty}m_{B}, \quad m_{G}^0=\lim_{\beta\rightarrow \infty}m_{G}.$$ We also introduce the total mixed susceptibility   
\begin{eqnarray}
\chi^0=p\chi_{G}^0+(1-p)\chi_{B}^0
\label{def:chi0tot}
\end{eqnarray}
and the mixed memory overlap $$m^0=pm_G^0+(1-p)m_B^0.$$
With these definitions we are able calculate the zero temperature limit of the saddle point equations:
\begin{equation}
\begin{split}
\chi_{B}^0= \sqrt{\frac{2}{\pi\alpha}}(1-\chi^0)e^{-{x^2}{2}}
\end{split}
\end{equation}
where $x$ is defined in Eq. (\ref{def:x})
and
\begin{equation}
\begin{split}
\chi_{G}^0=  \sqrt{\frac{2}{\pi\alpha}}(1-\chi^0) \frac{1}{\sqrt{1+\frac{[m_G^0(1-\chi^0)]^2}{\alpha(\chi_G^0)^2}}}
\end{split}
\label{eq:chi0G}
\end{equation}
\noindent where we have made use of (\ref{eq:lb}) and of the known limit (\ref{eqn:limite}). For the overlap with the bimodal pattern variables we  get the following
\begin{equation}
\begin{split}
&m_{B}^0= \mbox{Erf}\left(\frac{m^0}{\sqrt{2\alpha}}(1-\chi^0)\right), 
\end{split}
\end{equation}
while for the Gaussian contribution to the overlap we have
\begin{equation}
\begin{split}
&m_{G}^0= \frac{\chi_G^0(1-p)m_B^0}{1-p\chi_G^0}.
\end{split}
\end{equation}
From the last equation we can see that Eq. (\ref{eq:chi0G}) corresponds to the first term in the rhs of Eq. (\ref{eqn:chi0}).

In Figs.~\ref{fig:mb} and \ref{fig:mg} plots of both $m_B^0$  and $m_G^0$ are shown as functions of $p$ and for different value of the storage variable $\alpha$. It is worth noting that both the Gaussian and bimodal pattern overlaps maintain their highest possible values for every value of $p$ although they drop rapidly approaching the spinodal point of the phase diagram.

\section{Numerical analysis}
\label{section:III}

In this section we want to describe a numerical analysis, at zero temperature, of the mixed model that has the aim to find out the basin of attraction of the retrieval states, i.e. the size of the region of network states around each memory within which all states are attracted by the dynamical process to a close neighborhood of a memory, and the accuracy with which such states are retrieved in the phase diagram where these retrieval states exist at least as metastable. Then we calculate, through such simulations, the spinodal points at zero temperature. Since the results we have derived so far rely on replica symmetry and infinite size approximations, our numerical analysis will be affected by the violations of these two assumptions. In fact, replica symmetry breaking occurs nearby $\alpha_c$ at $T=0$ ~\cite{Amit:article}, while finite size effects appear in every computer simulation and cause the presence of spurious attractors at higher energy into which our zero temperature simulations can get stuck. Moreover, an artificial continuous Gaussian distribution for the memory variables will always be  discrete yielding further finite size effects.

\begin{figure}[t!]
\centering
\includegraphics[width=0.99\columnwidth]{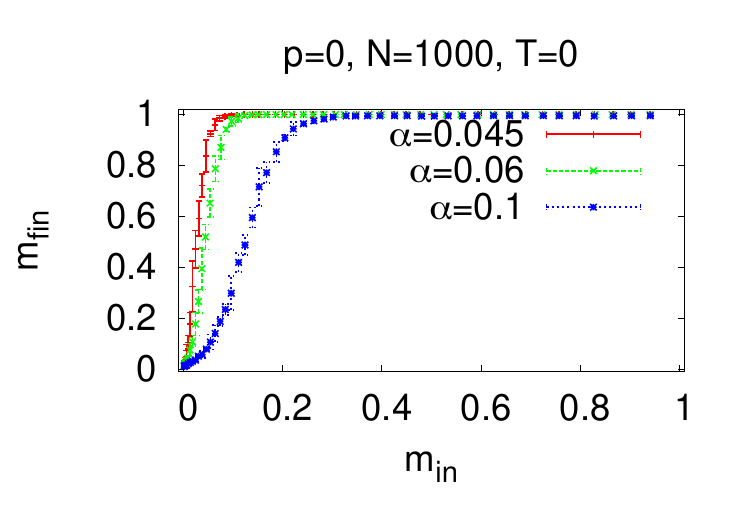} 
\caption{Final overlap with the pattern close to which we started the simulation as a function of the initial overlap with such pattern for three values of $\alpha$. The values used for $\alpha$ are $\alpha=0.045$ (red), $\alpha=0.06$ (green), $\alpha=0.1$ (blue).}
\label{fig:esempio}
\end{figure}

\begin{figure}[t!]
\includegraphics[width=0.99\columnwidth]{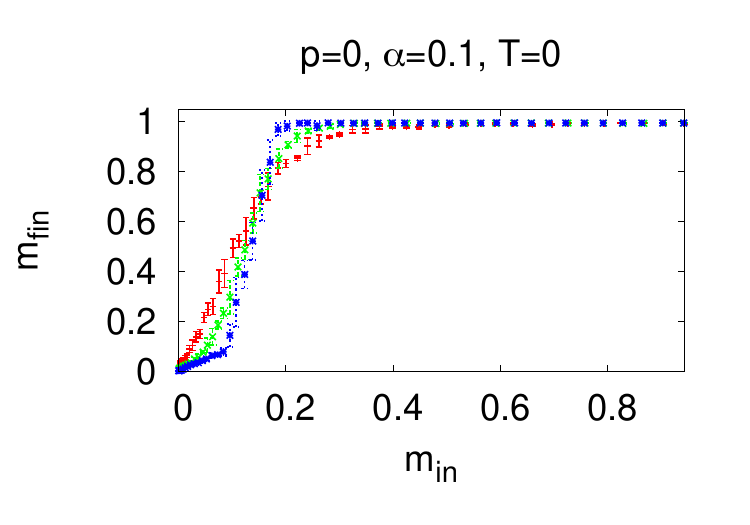}
\caption{Final overlap with the pattern aligned with the initial configuration as a function of the initial overlap with such pattern for three values of $N$. The used values of $N$ are $N=300$ (red), $N=1000$ (green), $N=3000$ (blue).}
\label{fig:esempio2}
\end{figure}

\begin{figure}[t!]
\centering
\includegraphics[width=0.99\columnwidth]{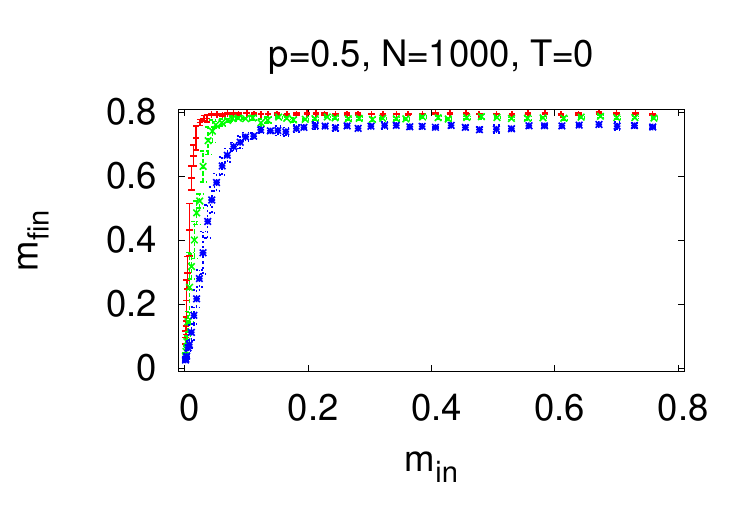} 
\caption{Final overlap with the pattern close to which we started the simulation as a function of the initial overlap with such pattern for three values of $\alpha$. The values used for $\alpha$ are $\alpha=0.01$ (red), $\alpha=0.02$ (green), $\alpha=0.03$ (blue).}
\label{fig:esempiop05}
\end{figure}

\begin{figure}[t!]
\includegraphics[width=0.99\columnwidth]{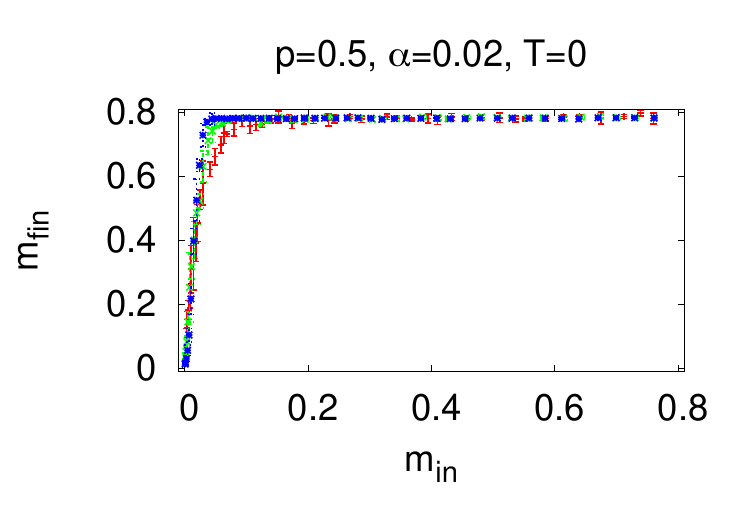}
\caption{Final overlap with the pattern aligned with the initial configuration as a function of the initial overlap with such pattern for three values of $N$. The used values of $N$ are $N=300$ (red), $N=1000$ (green), $N=3000$ (blue).}
\label{fig:esempio2p05}
\end{figure}

In order to consider all this features, we performed simulations using the so called Metropolis algorithm at $T=0$. First we comment the numerical results obtained in the region of the phase diagram that is known to be inside the theoretical MR spinodal line at $T=0$, Eq.~(\ref{eq:spine}), i.e., for $\alpha< 0.1379(1-p)^2$. 

In order to find the basins of attraction of the retrieval states, the simulation is started with an initial configuration that has different values of the overlap with one of the stored patterns. 
Eventually, the final overlap with such pattern is measured and it is averaged over a large number of samples. 
An example of what can be obtained with this analysis is shown in Fig.~\ref{fig:esempio} and ~\ref{fig:esempiop05}  with a plot of the final average overlap with one of the memories as a function of the initial overlap with this same memory. 
For all three values of $\alpha$ a common behavior can be noticed: for some values of the initial overlap, $m_{in}$, the final overlap is always the highest possible, i.e., the system falls in the retrieval states.
Besides, it is clear from Fig.~\ref{fig:esempio2} and ~\ref{fig:esempio2p05} that as $N$ grows the curves become sharper. 

\begin{figure}[t!]
\centering
\centering
\includegraphics[width=0.99\columnwidth]{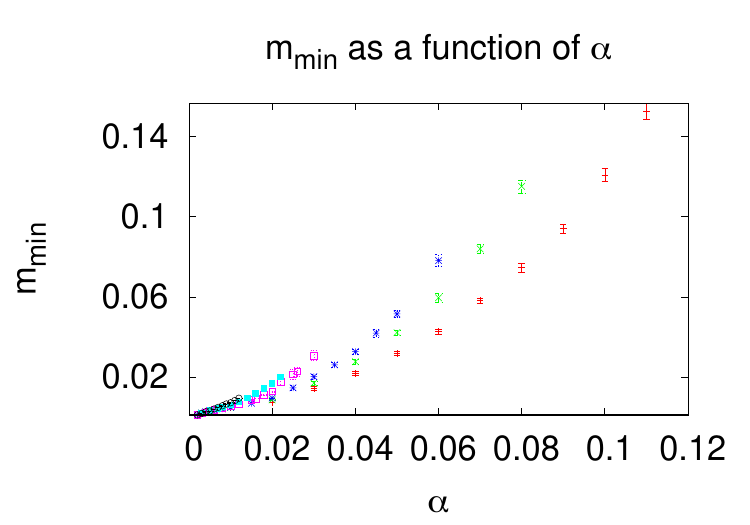}
\caption{The minimal initial pattern overlap $m_{min}$ is plotted as a function of $\alpha$ and for different values of $p$. Here $N=1000$ is the system size. The used values of $p$ are $p=0$ (red), $p=0.2$ (green), $p=0.3$ (blue), $p=0.5$ (purple),  $p=0.6$ (cyan),  $p=0.7$ (black).}
\label{fig:BS1}
\end{figure}

In our simulations we choose the operative definition of $m_{min}$ as the minimal initial overlap leading to a final overlap which is at least half the plateau value.
With this procedure we find the following results.
In Fig.~\ref{fig:BS1},  $m_{min}$ is plotted as a function of $\alpha$ and for different values of the variable $p$. Here we learn that the basin of attraction gets smaller and smaller as  $\alpha$  grows and the reason may be attributed to the fact that more patterns are added to the landscape and it is necessary to start closer to the retrieved memory in order to fall into it. 

\begin{figure}[t!]
\centering
\includegraphics[width=0.99\columnwidth]{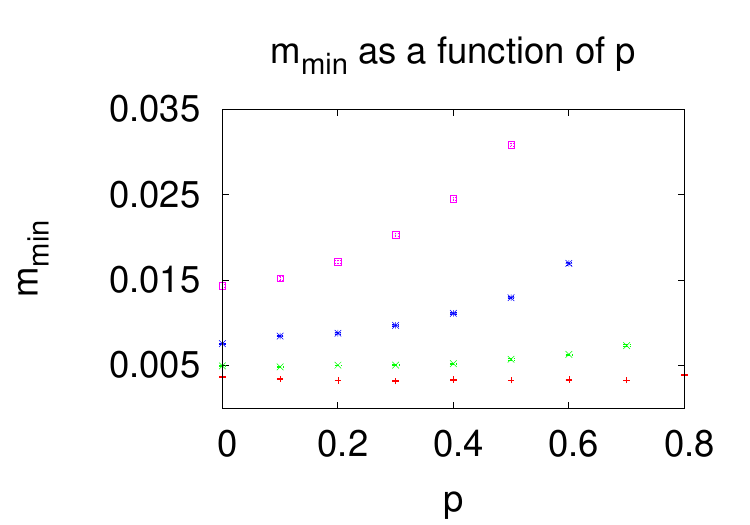}
\caption{$m_{min}$ is plotted as a function of $p$ and for different values of $\alpha$. Here $N=1000$ is the system size. $\alpha$ grows from bottom to top. The values used for $\alpha$ are $\alpha=0.005$ (red), $\alpha=0.01$ (green), $\alpha=0.02$ (blue), $\alpha=0.03$ (purple).}
\label{fig:BS3}
\end{figure}

This behavior is common for every value of $p$, but, as shown in Fig.~\ref{fig:BS3}, for small values of $\alpha$, $m_{min}$, as a first approximation is independent of $p$, which is not the case when $\alpha$ is sufficiently high. This means that, as long as we are in the retrieval phase, at a sufficiently small  $\alpha$ the system does not care if we are adding more Gaussian variables to the memories and the basin of attraction remains unchanged until we add too many for that fixed $\alpha$ and the basin of attraction starts to get smaller. 

\begin{figure}[t!]\centering
\includegraphics[width=0.99\columnwidth]{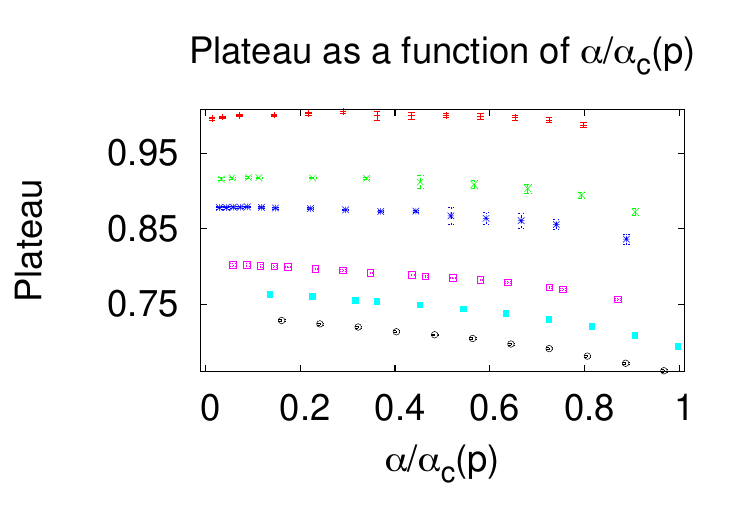}
\caption{The plateau values are plotted as functions of $\alpha/\alpha_c(p)$ and for different values of $p$. Here $N=1000$ is the system size. $p$ grows from top to bottom. The used values of $p$ are $p=0$ (red), $p=0.2$ (green), $p=0.3$ (blue), $p=0.5$ (purple),  $p=0.6$ (cyan),  $p=0.7$ (black).}
\label{fig:BS2}
\end{figure}

\begin{figure}[t!]
\centering
\includegraphics[width=0.99\columnwidth]{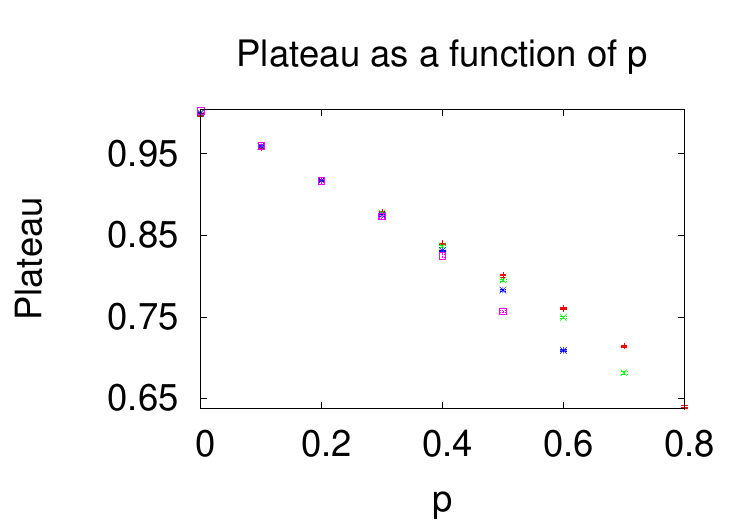}
\caption{The plateau values are plotted as functions of $p$ and for different values of $\alpha$. Here $N=1000$ is the system size. The values used for $\alpha$ are $\alpha=0.005$ (red), $\alpha=0.01$ (green), $\alpha=0.02$ (blue), $\alpha=0.03$ (purple).}
\label{fig:BS4}
\end{figure}

In Fig.~\ref{fig:BS2} the plateau, reached by the curves in Fig.~\ref{fig:esempio}, is plotted as a function of $\alpha$ and for different values of $p$. It is clear that the dependence on the memory loading variable becomes more apparent as $p$ grows, in agreement with what can be inferred by the plot in Fig.~\ref{fig:BS3}. Finally, the plateaus are linear functions of $p$ as can be seen in Fig.~\ref{fig:BS4}. 

Next we describe an analysis  to numerically estimate $\alpha_c(p)$ looking at the numerical behavior in the phase diagram region beyond the  theoretical spinodal line, i.e., where the thermodynamic limit theory predicts a spin-glass.
This capacity problem has already been a task of ~\cite{Amit:article}, but, there, the authors used a numerical approach finding $\alpha_c=0.145 \pm 0.01$ and conjecturing that this result were due to replica symmetry breaking. The conjecture seemed to be warranted by the 1RSB calculation done in  ~\cite{Crisanti:article} where it was found $\alpha_c^\text{1RSB}=0.144$, but a new result, in  ~\cite{Steffan:article, Steffan1:article}, corrected the former study giving $\alpha_c^\text{1RSB}=0.138$, thus proving the conjecture wrong, such result has been proved rigorously in ~\cite{Agliari2:article}. Furthermore, studies on the capacity problem have been presented in  ~\cite{capacity1:article} where $\alpha_c=0.1455\pm0.001$, in ~\cite{capacity2:article} with the numerical result $\alpha_c=0.143\pm0.001$ and in ~\cite{capacity3:article} with the analytical result $\alpha_c=0.159$ found in the RSB scheme of De Dominicis et al. ~\cite{Dominicis:article} .

At last, a correction on the approach of Ref.~\cite{Amit:article} was made by ~\cite{Stiefvater:article} giving $\alpha_c=0.141 \pm 0.0015$. 
This is the method that we will follow in the remaining analysis and it gives a storage capacity that is greater than the one found in the RS analysis. Besides, although closer to  the value of Ref.~\cite{Steffan:article}, it is still significantly higher, thus it cannot all be a consequence of replica symmetry breaking. A possible explanation can be found in finite size effects that, as stated before, give rise to attractors that are not seen by the analytical infinite size theory. The simulated dynamics can get trapped into these one-spin-flip stable metastable states from which they cannot escape at $T=0$. 

Following Ref.~\cite{Stiefvater:article} we start our Monte Carlo simulations at zero temperature with a configuration as close as possible to one of the memories. We are interested to the final overlap with the same memory. For each value of $\alpha$, $N$ and $p$, we repeat $100$ times the dynamics to obtain histograms of the final overlaps $m$.  

It has already been noticed in \cite{Amit:article} and \cite{Stiefvater:article}  that these distributions contain two peaks even above $\alpha_c$: a high-$m$ peak and a low-$m$ peak. The low-$m$ peak, sitting at about $m \simeq 0.35$, is considered a remnant magnetization, similar to that encountered in spin glasses \cite{Amit:book,Winzel:article} and not predicted by the replica symmetric theory. It is a finite size effect due to the presence, nearby the initial configuration, of a high number of attractors made of a combination of more than one pattern. 

\begin{figure}[t!]
\centering
\includegraphics[width=0.99\columnwidth]{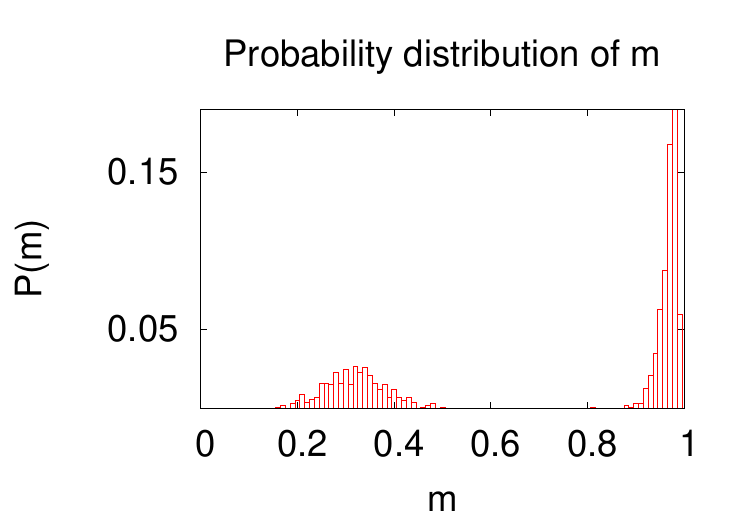}
\caption{Probability distribution of the order parameter $m$ above the theoretical $\alpha_c$. Here the parameters are $p=0$, $N=3000$, $\alpha=0.15$.}
\label{fig:dist3000}
\end{figure}

\begin{figure}[t!]
\centering
\includegraphics[width=0.99\columnwidth]{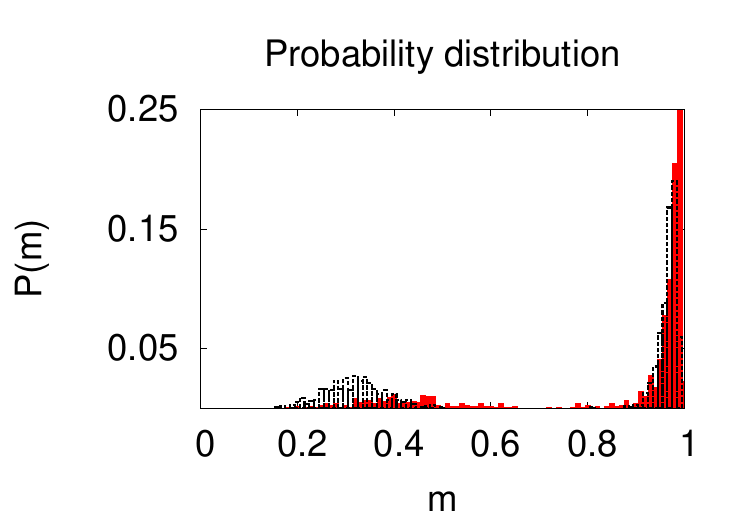}
\caption{Probability distribution of the order parameter for $p=0$ and $\alpha=0.15>\alpha_c$ and two values of the system size $N=1000$ (red) and $N=3000$ (black). The high-$m$ peak becomes smaller as $N$ grows.}
\label{fig:2dist}
\end{figure}

Actually, since we are beyond the  theoretical retrieval spinodal point, we would not expect to have the high-$m$ peak, as well. It is there because the finite size $\alpha_c$ will be higher than the theoretical one. An example of this distribution is given in Fig.~\ref{fig:dist3000}. 
From Fig.~\ref{fig:2dist}, we can see that a first order phase transition is occurring in the original Hopfield model $p=0$ because the high-$m$ peak lowers as $N$ grows. 

\begin{figure}[t!]
\centering
\includegraphics[width=0.99\columnwidth]{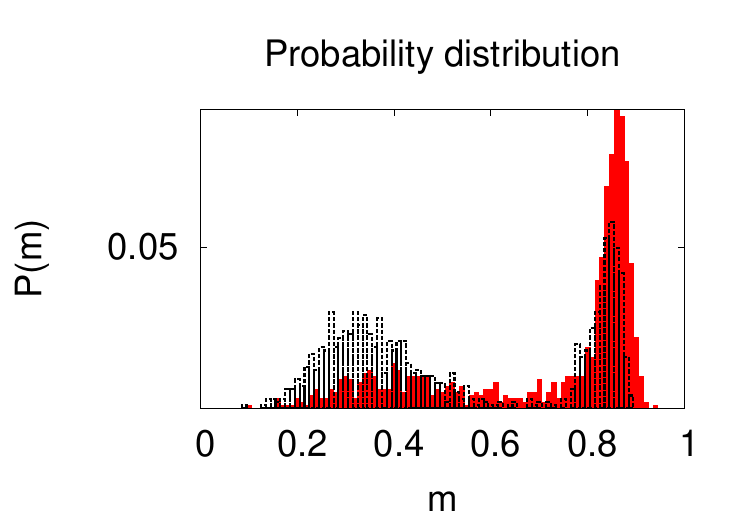}
\caption{Probability distribution of the order parameter for $p=0.5$ and $\alpha=0.05>\alpha_c(p)$ and two values of the system size $N=1000$ (red) and $N=3000$ (black). The high-$m$ peak becomes smaller as $N$ grows.}
\label{fig:2distp}
\end{figure}

This first order transition also occurs in the mixed case ($p>0$) as showed in Fig.~\ref{fig:2distp}. It may be useful to add that in the mixed case the remnant magnetization does not change much, while the high-$m$ peak moves toward smaller values linearly in $p$, as  analytically predicted.  


Assuming standard finite size scaling for first order transitions, we can write
\begin{equation}
f_N=\exp(A_N-B_N N), 
\end{equation}
where $f_N$ is the frequency with which the high peak is selected. The coefficients $A_N$ and $B_N$ are expected to be of $O(1)$ in the large $N$ limit. 
We also assume that they are self-averaging quantities and their mean values close to the critical point can be written as follows
\begin{eqnarray}
\left<A_N\right>&=&a_N
\nonumber
\\
\left<B_N\right>&=&b_N (\alpha-\alpha_c)+O((\alpha-\alpha_c)^2),
\nonumber
\end{eqnarray}
as $\alpha\rightarrow \alpha_c$ from the right, with $a_N$ and $b_N$ constants that approach $a$ and $b$ respectively as $N\rightarrow\infty$. 

Since $f_N$ is an exponential in $N$ it has large fluctuations and it cannot be self averaging. As a consequence if we want to average the quantity $\log(f)$ over the disorder, we have to perform a quenched average. This is done by constructing more than one histogram, by calculating $\log(f)$ for each one of them and, finally, by averaging it over the accumulated samples. 

\begin{figure}[t!]
\centering
\includegraphics[width=0.99\columnwidth]{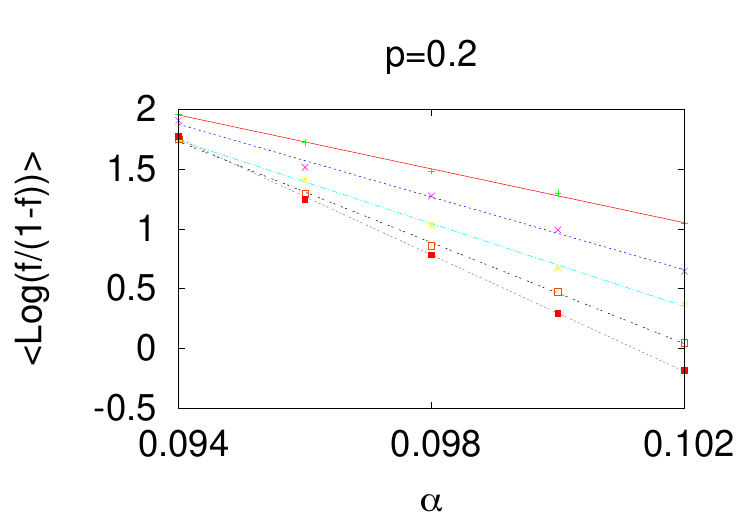}
\caption{Checking that the linear hypothesis for the finite size scaling function is a correct assumption at $p=0.2$. It has also been verified that the slope of these lines grows linearly with $N$. These lines are calculated at  $N=1000$ (red), $N=2000$ (blue), $N=3000$ (cyan), $N=4000$ (black), $N=5000$ (grey).}
\label{fig:p0}
\end{figure}

\begin{figure}[t!]
\centering
\includegraphics[width=0.99\columnwidth]{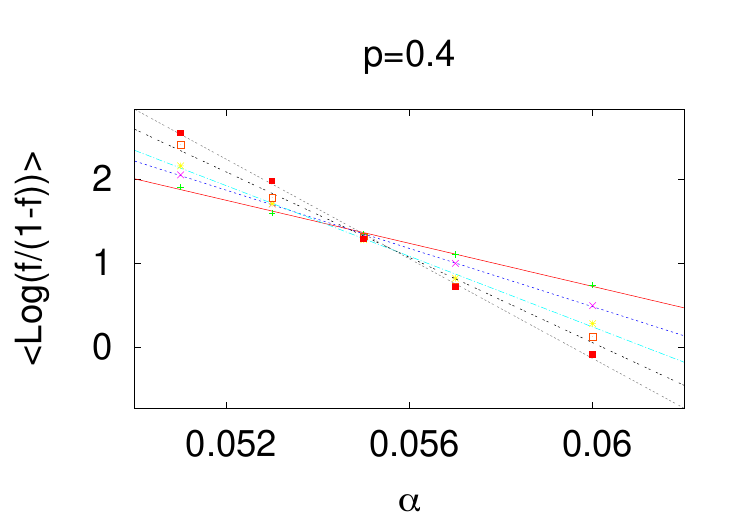}
\caption{Checking that the linear hypothesis for the finite size scaling function is a correct assumption at $p=0.4$. It has also been verified that the slope of these lines grows linearly with $N$. These lines are calculated at  $N=1000$ (red), $N=2000$ (blue), $N=3000$ (cyan), $N=4000$ (black), $N=5000$ (grey).}
\label{fig:p04}
\end{figure}

The data collected in the present study consists in histograms each constructed by $100$ different runs for every $N$ and $\alpha$, then for $N=1000$ the quenched average is performed over $200$ histograms, for $N=2000$ over 120 histograms and for $N=3000,4000,5000$ over $60$ histograms.
Moreover, as noticed in ~\cite{Stiefvater:article}, if one takes into account the quantity $$\log\left(\frac{f}{1-f}\right),$$ instead of $\log(f)$, higher orders terms in $\alpha-\alpha_c$ can be neglected and the form of the finite size scaling can be taken as:
\begin{equation}
\log\left(\frac{f}{1-f}\right)=a-b(\alpha-\alpha_c)N\;,
\label{eq:FSS}
\end{equation}
for $\alpha>\alpha_c$. Here  we have neglected higher order terms in $1/N$ for $a_N\simeq a$ and $b_N\simeq b$. The linearity property can be checked out each time, as examples see Figs.~\ref{fig:p0} and \ref{fig:p04} for two values of $p$. 
\begin{figure}[t!]
\centering
\includegraphics[width=0.99\columnwidth]{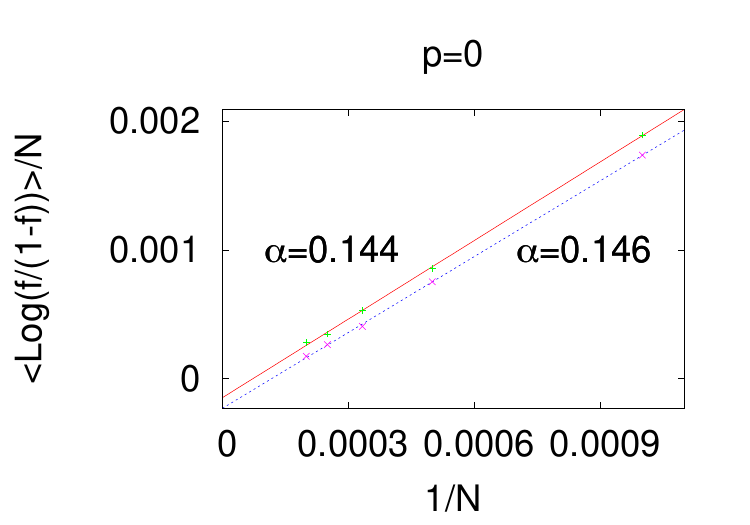}
\caption{Numerical analysis that allows to find the value of $\alpha_c$ by the knowledge of the intercepts of the two lines above calculated at the two values of $\alpha$ indicated. At $p=0$ we find $\alpha_c=0.1404\pm 0.0010$.}
\label{fig:analisip0}
\end{figure}

\begin{figure}[t!]
\centering
\includegraphics[width=0.99\columnwidth]{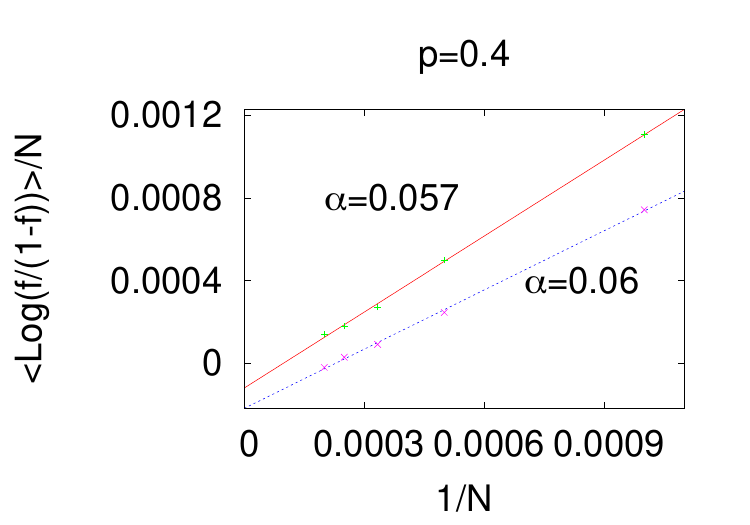}
\caption{Numerical analysis that allows to find the value of $\alpha_c$ by the knowledge of the intercepts of the two lines above calculated at the two values of $\alpha$ indicated. At $p=0.4$ we find $\alpha_c=0.0534\pm 0.0008$.}
\label{fig:analisip04}
\end{figure}

\begin{figure}[t!]
\centering
\includegraphics[width=0.99\columnwidth]{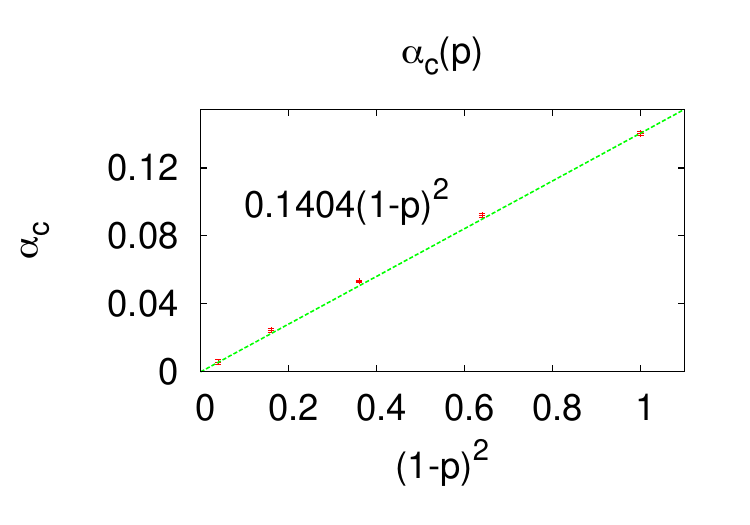}
\caption{Numerical results (red points) obtained for $\alpha_c$ as a function of $p$, it can be noticed that it follows a quadratic behavior just like in the analytical results (green curve).}
\label{fig:alphac1}
\end{figure}

The analysis proceeds by extrapolating from the graphs, like in Figs.~\ref{fig:analisip0} and \ref{fig:analisip04}, the intercepts of the two straight lines for two values of $\alpha$. This, together with Eq.~(\ref{eq:FSS}), will be enough to calculate $\alpha_c$. Here we give a few values obtained for some $p$:
\begin{equation}
\begin{split}
  p=0 &\quad \rightarrow \quad \alpha_c=0.1404\pm 0.0010 \\
p=0.2 &\quad \rightarrow \quad \alpha_c=0.0922\pm 0.0014\\
p=0.4 &\quad \rightarrow \quad \alpha_c=0.0534\pm 0.0008\\
p=0.6 &\quad \rightarrow \quad \alpha_c=0.0246\pm 0.0011\\
p=0.8 &\quad \rightarrow \quad \alpha_c=0.0058\pm 0.0015\\
\end{split}
\end{equation}
\noindent
that can be compared with those predicted by the analytical replica symmetric theory and given by Eq.~(\ref{eq:spine})
\begin{equation}
\begin{split}
  p=0 &\quad \rightarrow \quad \alpha_c=0.1379\\
p=0.2 &\quad \rightarrow \quad \alpha_c=0.0882\\
p=0.4 &\quad \rightarrow \quad \alpha_c=0.0496\\
p=0.6 &\quad \rightarrow \quad \alpha_c=0.0221\\
p=0.8 &\quad \rightarrow \quad \alpha_c=0.0055\\
\end{split}
\end{equation}
It is clear that the numerical results for $\alpha_c$ are larger than the replica symmetric analytical prediction.
This fact was observed in the standard Hopfield model and attributed to both replica symmetry breaking and finite size effects that add attractors, not predicted by the theory. 
Numerical data for $\alpha_c(p)$ are shown in Fig.~\ref{fig:alphac1}, confirming the quadratic behavior in $(1-p)$ as derived analytically in Eq.~(\ref{eq:spine}) by the RS theory.

\section{Conclusions}
\label{section:IV}

To summarize, we would like to point out a few features of the mixed Hopfield model that are responsible for the shrinking of the retrieval phase as more Gaussian variables are added to the patterns.
First of all, we notice, by looking at Eq.~(\ref{eq:freeE}), that the ferromagnetic term of the Gaussian variables can be rearranged in a way that it multiplies a factor $(1-1/\chi_G)$. From Eq.~(\ref{eq:mg}), we find
$$
\chi_G=\frac{m_G}{pm_G+(1-p)m_B}\;,
$$
Indeed \ $\chi_G$ is the fraction of the Gaussian contribution to the overall overlap therefor when we only have Gaussian variables in the patterns, $\chi_G=1$ and the ferromagnetic term disappears. 

Besides, since the purely Gaussian model shows a perfect spherical symmetry of the free energy as a function of the pattern overlap $m^{(\nu)}$ and since this symmetry is enforced as we add more and more Gaussian variables, we can assume that the local susceptibility, $\chi_G$, gives a measure of the spherical symmetry of the model and that when it reaches unity, this symmetry becomes perfect, the free energy lacks the ferromagnetic term and, consequently, retrieval gets lost for $\alpha>0$. 

Another way we can look at the problem is via Eqs.~(\ref{eq:qb}) and (\ref{eq:qg}).
Here we notice that the local fields acting on the variables depend on whether they correspond to the binary or the Gaussian components of the pattern: in both cases, local fields are random variables with a variance proportional to $\alpha$, but the mean of these random fields are non null only for variables with binary patterns.
As a consequence, in the purely Gaussian model retrieval will not be possible unless $\alpha=0$.



When the memories are present in an extensive number, i.e.\ $\alpha>0$, the Gaussian overlap of the aligned patterns works as a spin glass noise added to that of the misaligned ones. We have demonstrated, both numerically and analytically, that the capacity at zero temperature drops in a quadratic way with the fraction $1-p$ of binary variables in the patterns, while at $T>0$ the critical lines can be perfectly fitted by a linear combination of $(1-p)^2$ and $(1-p)^4$ terms (that is, they are very smooth functions going to zero not slower than $(1-p)^2$).
By Monte Carlo simulations, we have measured the same quadratic behavior in the actual capacity $\alpha_c(p)$ which turns out to be slightly larger than the RS prediction.

Approaching the purely Gaussian model, that is $p \to 1$ , the critical lines get squeezed towards $\alpha=0$, but at the same time the critical point at $T_g=1$ remains unchanged.
So, even at $\alpha=0$ there is a retrieval phase, if one stores a sub-extensive number of memories. This sub-extensive retrieval phase has been studied in the context of immune networks \cite{ImmuneNetworks2013}, but not in the Hopfield model. We leave it as a future challenge, particularly interesting for developing the proper theoretical framework to experiments like those in \cite{Ruocco:article}.

In the present work we have characterized the retrieval states varying $p$, with a particular interest in checking what happens in the purely Gaussian limit.
We have found that the basin of attraction of the retrieval states, that is the minimal overlap needed to retrieve a memorized pattern, mainly depends on $\alpha$ and has a very tiny dependence on $p$. This means that the few patterns one can store in the Gaussian limit can be still retrieved easily.
Moreover the retrieval accuracy does depend on $p$ (and very little on $\alpha$) but even in the $p\to1$ limit does not go to zero. So the stored patterns can be retrieved with a non zero accuracy even in the Gaussian limit.

In conclusion, the picture that comes out from our observations is that even in the Gaussian limit where patterns are largely mismatched and only few of them can be stored in the Hopfield model, the retrieval phase is well behaved. By this we mean that memorized patterns can be easily retrieved (large basin of attraction) although with a non perfect accuracy (due to the extra noise induced by the mismatched variables in the pattern).

We believe it is very interesting to deepen the study of the retrieval phase in the $\alpha=0$ limit, as modern artificial neural networks works in the largely over-parameterized regime where the number of neurons is much larger (by order of magnitude) than the number of classes they need to store.


\appendix

\section{Free energy of the MHM}
\label{app:A}

First, we calculate the replicated average partition function  separating the contribution of the Gaussian $\xi_{i,G}$ and of the bimodal $\xi_{i,B}$ variables of the memories:
\begin{eqnarray}
\label{eq:Zn}
\mathbb{E}[Z^n]=
e^{\frac{-\beta \alpha N n}{2}}\sum_{\{s_\rho\}}\mathbb{E}\left[f_{C}(\xi,s^\rho)\, f_{NC}(\xi,s^\rho)\right]
\end{eqnarray}
where we have defined 
\begin{eqnarray}
\label{eq:fC}
f_{C}(\xi,s^\rho)&\equiv&
\beta\sum_{\nu=1}^s h^{(\nu)} \sum_{\rho=1}^n\Bigg(\sum_{i=0}^{pN-1}\xi_{i,G}^{(\nu)} s_{i}^\rho + \sum_{i=pN}^{N}\xi_{i,B}^{(\nu)} s_{i}^\rho \Bigg)\\
\nonumber
f_{NC}(\xi,s^\rho)&\equiv& 
 \int \prod_{\mu=1}^{\alpha N} \prod_{\rho=1}^n dm_{\rho}^{(\mu)}
 \sqrt{\frac{\beta N}{2\pi}} 
 \exp\Biggl\{ - \frac{\beta N}{2}\sum_{\mu=1}^{\alpha N}\sum_{ \rho=1}^n(m_{\rho}^{(\mu)})^2  \\
 \label{eq:fNC}
&& +\beta \sum_{\mu=1}^{\alpha N} \sum_{\rho=1}^n m_{\rho}^{(\mu)} \Bigg(\sum_{i=0}^{pN-1}\xi_{i,G}^{(\mu)} s_{i}^\rho +\sum_{i=pN}^{N}\xi_{i,B}^{(\mu)} s_{i}^\rho \Bigg)\Bigg\}
\end{eqnarray}
as the contribution of the aligned and misaligned patterns respectively. 
If we perform the average over the disorder for the Gaussian elements of the misaligned patterns in (\ref{eq:fNC}) we get for Gaussian and bimodal elements
\begin{multline}
\mathbb{E}_G\Bigg[\exp\left(\beta \sum_{\mu \rho}m_{\rho}^{(\mu)}\sum_{i=0}^{pN-1}\xi_{i,G}^{(\mu)} s_{i}^\rho\right)\Bigg]\\
=\exp\left\{\frac{\beta^2}{2} \sum_{\mu \rho \sigma}\sum_{i=0}^{pN-1} m_{\rho}^{(\mu)} m_{\sigma}^{(\mu)} s_{i}^\rho s_{i}^\sigma\right\}
\end{multline}
\begin{multline}
\mathbb{E}_B\Bigg[\exp\left(\beta \sum_{\mu \rho}m_{\rho}^{(\mu)} \sum_{i=pN}^{N}\xi_{i,B}^{(\mu)} s_{i}^\rho\right)\Bigg]=\\
\prod_{i=pN}^{N}\prod_{\mu=1}^{\alpha N} \cosh \left(\beta \sum_{\rho=1}^n m_{\rho}^{(\mu)} s_i \right)
\end{multline}
We, then,  rescale $m^{(\mu)}_{\rho}\rightarrow \frac{m^{(\mu)}_{\rho}}{\sqrt{\beta N}} $   in order to get the right thermodynamic limit.  Since in this limit the number of misaligned patterns diverges, we can approximate the average over the bimodal distribution with a Gaussian average, as well.  In this sense the term of the misaligned memories is exactly equal to the one of the original Hopfield model. As a consequence, the transition line $T_g(\alpha)$ between the paramagnetic phase and the spin glass phase does not change in the mixed case.
 
Considering only the partition function of the misaligned patterns, we further proceed by reassembling the terms and, by means of a Dirac's delta, we introduce the overlap between configurations, $$q_{\rho\sigma} \equiv \frac{1}{N}\sum_{i=1}^N s_i^{\rho}s_i^{\sigma}$$ that is the order parameter that detects the spin glass phase transition. Finally, we are left with the following expression for the contribution of the misaligned patterns
\begin{multline}
\mathbb{E}\left[f_{NC}(\xi,s^\rho)\right]=
\int  \prod_{\rho<\sigma}^{1,n} dq_{\rho\sigma}\prod_{\rho<\sigma}^{1,n} dr_{\rho\sigma}  \\
\exp\left\{-\frac{\alpha N}{2}\mbox{Tr}  \ln\left[(1-\beta)\mathbb{I}   -\beta \hat{q} \right]\right\} \\
\exp\left\{ -\frac{\alpha \beta^2  N}{2}\sum_{\rho\neq\sigma}^{1,n} r_{\rho\sigma}q_{\rho\sigma}+\frac{\alpha \beta^2}{2}\sum_{i=1}^N\sum_{\rho\neq\sigma}^{1,n} r_{\rho\sigma}s_{i}^\rho s_{i}^\sigma    \right\}
\label{eq:EfNC1}
\end{multline}
where $\mathbb{I}$ is the identity matrix and where we have used the Laplace transform expression of the Dirac's delta with $r_{\rho\sigma}$  Lagrange multipliers. The independent overlaps and multipliers are for $\rho<\sigma$. By construction $q_{ab}=q_{ba}$, $r_{ab}=r_{ba}$ and the diagonal elements are zero.  We have neglected some proportionality constants that multiply the measure and do not influence the outcome of the saddle point equations. 

Now we can work out the aligned pattern partition function contribution (\ref{eq:fC}). We can exactly get the average over the Gaussian extracted patterns as
\begin{multline}
\mathbb{E}_G\left[\exp\left\{ \sum_{\nu=1}^s\sum_{\rho=1}^N \beta  (m_\rho^{(\nu)}+h^{(\nu)})\sum_{i=0}^{pN-1} \xi_{i,G}^{(\nu)} s_i^\rho \right\}\right]=\\
=\exp\left\{  \frac{\beta^2}{2}\sum_{\nu=}^s  \sum_{\rho\sigma}^{1,n}(m_\rho^{(\nu)}+h^{(\nu)})(m_\sigma^{(\nu)}+h^{(\nu)})\sum_{i=0}^{pN-1} s_i^\rho s_i^\sigma\right\}
\end{multline}
So that the total partition function, so far, reads

\begin{widetext}
\begin{multline}
\mathbb{E}[Z^n]=e^{\frac{-\beta \alpha N n}{2}} \int  \prod_{\rho<\sigma}^{1,n} dq_{\rho\sigma} dr_{\rho\sigma}  \prod_{\nu=1}^s \prod_{\rho=1}^n \frac{dm_{\rho}^{(\nu)}}{\sqrt{2\pi}}
\exp\biggl\{ -\frac{\beta N}{2}\sum_{\nu\rho}(m_{\rho}^{(\nu)})^2 -\frac{\alpha N}{2}\mbox{Tr} \ln\left[ (1-\beta)\mathbb{I}   -\beta \hat{q} \right] -\frac{\alpha \beta^2  N}{2}\sum_{\rho\neq\sigma}r_{\rho\sigma}q_{\rho\sigma}\biggr\} \\
\sum_{\{s_\rho\}} \prod_{i=1}^{pN-1} \exp\Bigg\{\frac{\beta^2}{2}\sum_{\rho\sigma}^{1,n}\sum_{\nu=1}^s (m_\rho^{(\nu)}+h^{(\nu)})(m_\sigma^{(\nu)}+h^{(\nu)}) s_i^\rho s_i^\sigma
+\frac{\alpha \beta^2}{2}\sum_{\rho\neq\sigma}r_{\rho\sigma}s_{i}^\rho s_{i}^\sigma \Bigg\}
\prod_{i=pN}^{N} \mathbb{E}_B
\Bigg[
\exp\left\{
\beta\sum_{\nu\rho}\left(m_\rho^{(\nu)}+h^{(\nu)}\right) s_i^\rho \xi_{i,B}^{(\nu)} + \frac{\alpha \beta^2}{2} \sum_{\rho\neq\sigma}^{1,n} r_{\rho\sigma}s_{i}^\rho s_{i}^\sigma 
\right\}\Bigg]\nonumber
\end{multline}
And, at last, we can write free energy per spin: $f=\lim_{n\to 0} f_n$ with 
\begin{multline}
f_n = \frac{\alpha}{2} +\frac{1}{2n}\sum_{\nu\rho}(m_{\rho}^{(\nu)})^2 +\frac{\alpha}{2\beta n} \mbox{Tr} \ln\left((1-\beta)\mathbb{I} -\beta \hat{q} \right) +\frac{\alpha \beta }{2n}\sum_{\rho\neq\sigma}r_{\rho\sigma}q_{\rho\sigma}\\
-\frac{p}{n\beta} \ln \Bigg[ \sum_{s_{\rho}}\exp\Bigg\{ \frac{\beta^2}{2}\sum_{\rho\sigma} \Bigg( \sum_\nu(m_\rho^{(\nu)}+h^{(\nu)})(m_\sigma^{(\nu)}+h^{(\nu)})
+\alpha r_{\rho\sigma}\Bigg) s^\rho s^\sigma \Bigg]\\
-\frac{(1-p)}{n\beta}\mathbb{E}_B\Bigg[\ln \sum_{s_{\rho}}\exp\Bigg\{ \beta\sum_{\nu=1}^s\sum_{\rho=1}^n(m_\rho^{(\nu)}+h^{(\nu)})\xi_{B}^{(\nu)} s^\rho+
\frac{\alpha \beta^2}{2}\sum_{\rho\neq\sigma}^{1,n}r_{\rho\sigma}s^\rho s^\sigma \Bigg\}\Bigg]
\end{multline}
\end{widetext}
The free energy in the replica symmetric case, given by Eq.~(\ref{eq:FE1}), is obtained by writing $q_{\rho\sigma}=(1-\delta_{\rho\sigma}) q$, $r_{\rho\sigma}=(1-\delta_{\rho\sigma}) r$ and $m^{(\nu)}_{\rho}=m^{(\nu)}$ and, the, by taking the limit $n\to 0$, as prescribed by the replica method.


\section{Memory retrieval-loss phase transition transition} 
\label{app:spineT}

At this point we define the functions:
 \begin{equation}
\begin{split}
&u_1(q,m)=m - (1-p)I_1-p\beta m (1-J_2) \\
&u_2(q,m)= q-(1-p)I_2-pJ_2
\end{split}
\end{equation}
The spinonal lines of the memory retrieval phase
are defined thorugh the following system of equations
\begin{equation}
\begin{sistema}
u_1(q,m)=0\\
u_2(q,m)=0\\
\frac{\partial u_1}{\partial m}\frac{\partial u_2}{\partial q}-\frac{\partial u_1}{\partial q}\frac{\partial u_2}{\partial m}=0
\label{eqn:sistemaspinodale}
\end{sistema}
\end{equation} 
The third equation has to be computed using the formulas listed below
\begin{equation}
\begin{split}
&\partial_m J_2 = 2m\beta^2(1-4J_2+3J_4) \\
&\partial_q J_2= \alpha \frac{T-1-q}{(T-1+q)^3}(1-4J_2+3J_4)\\
&\partial_m I_2= 2\beta(I_1-I_3)\\
&\partial_q I_2= \alpha \frac{T-1-q}{(T-1+q)^3}(1-4I_2+3I_4)\\
&\partial_m I_1=\beta(1-I_2)\\
&\partial_q I_1= \alpha \frac{T-1-q}{(T-1+q)^3}(I_3-I_1)
\end{split}
\end{equation} 
so that
\begin{equation}
\begin{split}
&\frac{\partial u_1}{\partial m}=1-\beta (1-I_2)-\beta p(I_2-J_2)+2p\beta^3m^2(1-4J_2+3J_4)\\
&\frac{\partial u_1}{\partial q}=\alpha \frac{T-1-q}{(T-1+q)^3}(I_1-I_3+\\
&+p(I_3-I_1+m\beta(1-4J_2+3J_4))) \\
&\frac{\partial u_2}{\partial m}=-2\beta(I_1-I_3)+2p\beta(I_1-I_3-m\beta(1-4J_2+3J_4))\\
&\frac{\partial u_2}{\partial q}=1+\alpha \frac{T-1-q}{(T-1+q)^3}(-(1-4I_2+3I_4)+\\
&+p(4(J_2-I_2)+3(I_4-J_4)))
\end{split}
\end{equation} 
The system in Eq.~(\ref{eqn:sistemaspinodale}) can be solved numerically and the spinodal lines for different values of $p$ can be drawn: the result will be an $\alpha_c(p,T)$.
To compute the first order transition line the system of equations  is 
\begin{eqnarray}
u_1(q,m)&=&0\\
u_2(q,m)&=&0\\
f(m(m,q),q(m,q))&=&f(0,q(0,q))
\label{eqn:sistemacritico}
\end{eqnarray} 
i.e.\ we are wondering where the free energy of the spin glass state (the free energy calculated at $m=0$ and at a $q$ that satisfies the saddle point equation with $m=0$ ) is equal to the free energy of the retrieval state (the free energy calculated at $m$ and $q$ that satisfy the system of saddle point equations). Again, system \ref{eqn:sistemacritico} can be solved numerically and its results are shown in Fig.~\ref{fig:alphat}.

\section{Coefficients of spinodal ad transition lines} 
\label{app:C1}

\begin{figure}[t!]
\includegraphics[width=0.99\columnwidth]{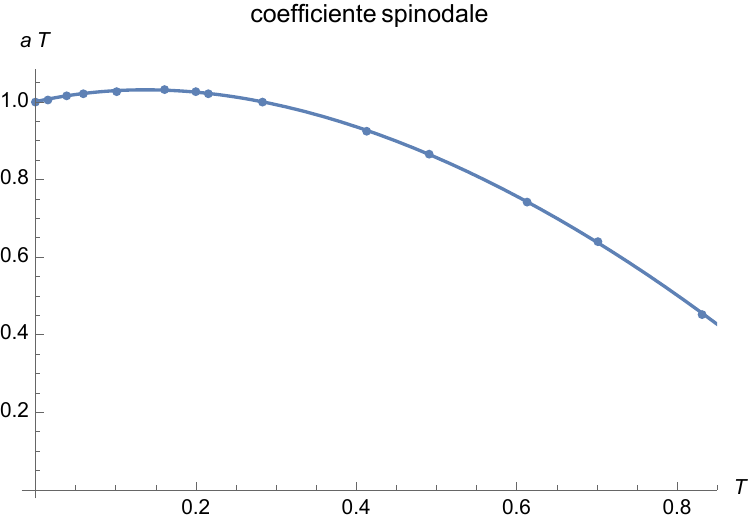}
\caption{Plot of of the interpolation of $a(T)$. The fit yields: $a(T)=1.000(1)+ 0.48(2)T - 2.0(1)T^2 + 1.1(1) T^3 - 0.5(1) T^4$.}
\label{fig:coeff_spinodale}
\end{figure}

In order to draw the spinodal lines we can fit the curves showed in \ref{fig:alphaspinodale} with the following function
\begin{equation}
\begin{split}
\label{eq:fit}
&\frac{\alpha_c(p,T)}{\alpha_c(0,T)}=a(T)(1-p)^2
+(1-a(T))(1-p)^4
\end{split}
\end{equation}
so that we can find the coefficient $a(T)$ for different values of the temperature. A second fit for $a(T)$, shown in Fig.~\ref{fig:coeff_spinodale}, provides
\begin{equation}
\begin{split}
&a(T)= 1.000(1)+ 0.48(2)T - 2.0(1)T^2 +\\
&+ 1.1(1) T^3 - 0.5(1) T^4.
\end{split}
\end{equation}
It is worth noticing that the function \ref{eq:fit} is only an approximation due to the fact that we have neglected higher order terms and, as a consequence, the coefficient $a(T)$ has a small systematic error because the function \ref{eq:fit} cannot fit the data perfectly. The expected limit $a(0)=1$ is recovered. In Fig. \ref{fig:coeff_spinodale} we notice that for an interval of low temperatures $a(T)>1$, which means that some of the lines in plot \ref{fig:alphaspinodale}, for this interval of temperatures, lie above the bisector. 

At last, from the spinodal $\alpha_c(0,T)$ line of the standard Hopfield model, we have those of the mixed model for every  $p$.

\begin{figure}[t]
\includegraphics[width=0.99\columnwidth]{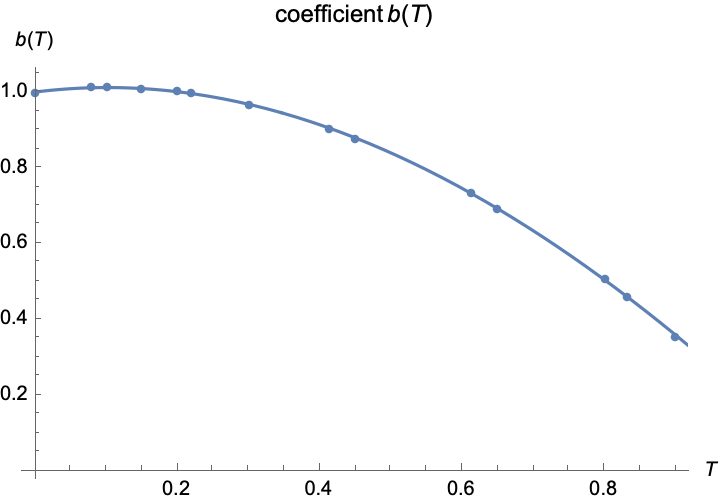}
\caption{Fit of the coefficient $b(T)=1.000(2)+ 0.29(2) T - 1.5 (1)T^2 + 0.7(1) T^3 - 0.3(1)T^4$.}
\label{fig:coeffcritico}
\end{figure}

\begin{figure}[t]
\centering
\includegraphics[width=0.99\columnwidth]{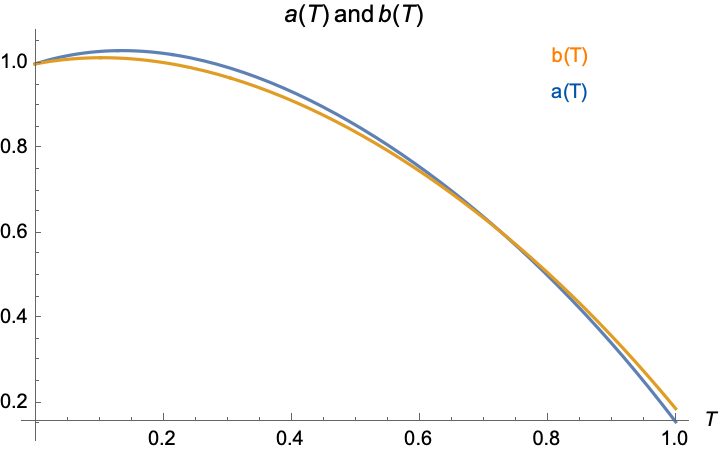}
\caption{Coefficients $b(T)$ and $a(T)$ compared.}
\label{fig:coeffs}
\end{figure}

For the transition lines we can carry out the same analysis as above, i.e. we can fit such curves shown in Fig.~\ref{fig:alphat} with the function $$\frac{\alpha_t(p)}{\alpha_t(0)}=b(T)(1-p)^2+(1-b(T))(1-p)^4$$ and, after finding a sufficient number of values of $b(T)$, we can in turn find a fit for this last coefficient (Fig.~\ref{fig:coeffcritico}):
\begin{equation}
\label{eq:fit2}
\begin{split}
&b(T)=1.000(2)+ 0.29(2) T - 1.5 (1)T^2 + \\
&+0.7(1) T^3 - 0.3(1)T^4
\end{split}
\end{equation}
Again there is a systematic error in the coefficient $b(T)$ due to the neglecting of higher order terms in the fitting function \ref{eq:fit2}. Since we have $b(T=0)=1$ we are able to observe that at zero temperature  $\alpha_t(p,0)=\alpha_t(0,0)(1-p)^2$ just like in the spinodal case. Also for the first order transition line we find an interval of temperatures in which $b(T)>1$, i.e., some of the curves in plot  \ref{fig:alphat} lie above the bisector. 
It may be of interest, to say that although the way the spinodal and transition lines depend on $(1-p)^2$ may be very similar, it is not exactly equal. In fact this dependence is enclosed in the coefficients $a(T)$ and $b(T)$, that can be seen plotted together in Fig.~\ref{fig:coeffs}.

Knowing the value of $\alpha_t(0,T)$, i.e., the transition line for the standard Hopfield model it is possible to draw the transition line for every value of $p$, as it is done in Fig.~\ref{fig:transizione}.

\section{Free energy of the MHM with separated variables}
\label{app:C}

In this appendix, we will calculate the free energy of the mixed model by separating the Gaussian and bimodal contributions of the patters.
To this end, we can rewrite the partition function as follows
\begin{equation}
\mathbb{E}[Z^n]=e^{\frac{-\beta \alpha N n}{2}}\sum_{\{s_\rho\}}\mathbb{E}\left[f_{C}(\xi,s^\rho)\, f_{NC}(\xi,s^\rho)\right]
\label{eq:Zmixed}
\end{equation}
where $f_C$ and $f_{NC}$ are defined in Eqs. (\ref{eq:fC})-(\ref{eq:fNC}) and we have already separated the aligned patterns, whose index is identified as $\nu=1,\ldots,s$, from the misaligned patterns, whose index is identified as $\mu=1,\ldots,P=\alpha N$. Next, we take into account only the misaligned patterns and we perform the same calculations we have performed in \ref{subsection:IIA}, Eq. (\ref{eq:EfNC1}), but this time we introduce two delta functions in order to have a Gaussian configuration overlap ($q^G$) and a bimodal one ($q^B$). Then, as we have already done, we compute the Gaussian integral in the variable $m$ and we introduce the Lagrangian multipliers for the two delta functions. Such variables can be indicated as $r^G$ and $r^B$, with a natural choice of notation. In this way the contribution to the partition function of the misaligned patterns reads as follows
\begin{widetext}
\begin{multline}
\label{eq:EfNC}
\mathbb{E}\left[f_{NC}(\xi, s^\rho)\right]=\int  \prod_{\rho<\sigma}^{1,n} dq_{\rho\sigma}^G dr_{\rho\sigma}^G dq_{\rho\sigma}^B dr_{\rho\sigma}^B \exp\left\{-\frac{\alpha N}{2}\mbox{Tr } \ln\left[(1-\beta)\mathbb{I}   -\beta \hat{q} \right]\right\}\\
\exp\left\{ -\frac{p\alpha \beta^2  N}{2}\sum_{\rho\neq\sigma}^{1,n} r_{\rho\sigma}^Gq_{\rho\sigma}^G -\frac{(1-p)\alpha \beta^2 N}{2} \sum_{\rho\neq\sigma}^{1,n} r_{\rho\sigma}^B q_{\rho\sigma}^B
\frac{\alpha \beta^2}{2} \sum_{i=0}^{pN-1} \sum_{\rho\neq\sigma}^{1,n} r_{\rho\sigma}^G s_{i}^\rho s_{i}^\sigma + \frac{\alpha \beta^2}{2} \sum_{i=pN}^{N} \sum_{\rho\neq\sigma}^{1,n} r_{\rho\sigma}^Bs_{i}^\rho s_{i}^\sigma \right\}
\end{multline}
We, then, work out the partition function concerning the aligned patterns by introducing two Dirac's deltas for the two types of contributions to the overlap and by writing them in their exponential forms with Lagrangian multipliers $\lambda_G$ and $\lambda_B$. Sending the condensing fields $h^{(\nu)}$ to zero we obtain 
\begin{multline}
f_{C}(\xi, s^\rho)=\int \prod_{\rho\nu} \frac{dm_{\rho,B}^{(\nu)} dm_{\rho,G}^{(\nu)} }{2\pi} d\lambda_{\rho_G}^{(\nu)} d\lambda_{\rho_B}^{(\nu)}
\exp\Bigg\{ \frac{N\beta}{2}\sum_{\rho \nu}\left[pm_{\rho,G}^{(\nu)} +(1-p)m_{\rho,B}^{(\nu)} \right]^2\\
-N\beta p \sum_{\nu \rho}m_{\rho,G}^{(\nu)} \lambda_{\rho_G}^{(\nu)} +\beta\sum_{\nu \rho} \lambda_{\rho_G}^{(\nu)}\sum_{i=0}^{pN-1}\xi_{i,G}^{(\nu)} s_{i}^\rho
-N\beta (1-p)\sum_{\nu \rho}m_{\rho,B}^{(\nu)} \lambda_{\rho_B}^{(\nu)} +\beta\sum_{\nu \rho} \lambda_{\rho_B}^{(\nu)}\sum_{i=pN}^{N}\xi_{i,B}^{(\nu)} s_{i}^\rho\Bigg\}
\label{eq:47}
\end{multline}
The average over the disorder of the Gaussian variables contribution, $\xi_{i,G}^{(\nu)}$ turns out to be
\begin{equation}
\mathbb{E}_G\left[\exp\left\{\beta\sum_{\nu \rho} \lambda_{\rho_G}^{(\nu)}\sum_{i=0}^{pN-1}\xi_{i,G}^{(\nu)} s_{i}^\rho\right\}\right]=\exp\left\{\frac{pN\beta^2}{2}\sum_\nu\sum_{\rho\neq \sigma}  \lambda_{\rho_G}^{(\nu)} \lambda_{\sigma_G}^{(\nu)} q_{\rho\sigma}^G+\frac{pN\beta^2}{2}\sum_{\nu\rho}  (\lambda_{\rho_G}^{(\nu)})^2
\right\}
\label{eq:48}
\end{equation}
The saddle point equation (\ref{eq:Zmixed}) for the variable $\lambda_{\rho_G}^{(\nu)}$ leads to the expression in matrix notation
\begin{equation}
\hat{\lambda}^G=\frac{1}{\beta}(\hat{q}^G+\mathbb{I})^{-1}\hat{m}^G
\end{equation}
Substituting in Eqs. (\ref{eq:47})-(\ref{eq:48}), the average $f_C$ becomes
\begin{multline}
\label{eq:EfC}
\mathbb{E}[f_{C}(\xi, s^\rho)]=
\int \prod_{\rho\nu} \frac{dm_{\rho,B}^{(\nu)} dm_{\rho,G}^{(\nu)} }{2\pi} d\lambda_{\rho_B}^{(\nu)}
\exp\Bigg\{ \frac{N}{2}\Bigg[\beta \sum_{\rho \nu}\left[pm_{\rho,G}^{(\nu)} +(1-p)m_{\rho,B}^{(\nu)} \right]^2\\
-p\sum_{\nu}\sum_{\rho\sigma} m_{\rho,G}^{(\nu)}  (\hat{q}^G+\mathbb{I})_{\rho\sigma}^{-1} m_{\sigma,G}^{(\nu)}
 - 2 \beta (1-p)\sum_{\nu \rho}m_{\rho,B}^{(\nu)} \lambda_{\rho_B}^{(\nu)} \Bigg] \Bigg\} \times
\mathbb{E}_B\left[\exp\left\{\beta\sum_{\nu \rho} \lambda_{\rho_B}^{(\nu)}\sum_{i=pN}^{N}\xi_{i,B}^{(\nu)} s_{i}^\rho \right\} \right]
\end{multline}
Multiplying the two contributions (\ref{eq:EfNC}) and (\ref{eq:EfC}) we have
\begin{equation}
\nonumber
\mathbb{E}[Z^n]=\int \prod_{\rho=1}^n\prod_{\nu=1}^s \frac{dm_{\rho,B}^{(\nu)} dm_{\rho,G}^{(\nu)} }{2\pi} d\lambda_{\rho_B}^{(\nu)} \prod_{\rho<\sigma}^{1,n} dq_{\rho\sigma}^Gdr_{\rho\sigma}^Gdq_{\rho\sigma}^Bdr_{\rho\sigma}^B \exp\left\{NA\left[q^B,q^G,r^B,r^G,m_B,m_G\right] \right\}
\end{equation}
with 
\begin{multline}
A \equiv -\frac{p\alpha \beta^2  }{2}\sum_{\rho\neq\sigma}r_{\rho\sigma}^Gq_{\rho\sigma}^G 
-\frac{(1-p)\alpha \beta^2  }{2}\sum_{\rho\neq\sigma}r_{\rho\sigma}^Bq_{\rho\sigma}^B
+\frac{\beta}{2}\sum_{\rho=1}^n\sum_{ \nu=1}^s\left[p\, m_{\rho,G}^{(\nu)} +(1-p)m_{\rho,B}^{(\nu)} \right]^2
-\frac{\alpha}{2}\mbox{Tr } \ln\left[(1-\beta)\mathbb{I}   -\beta \hat{q} \right]\\
-\frac{p}{2}\sum_{\nu=1}^s\sum_{\rho\sigma}^{1,n} m_{\rho,G}^{(\nu)}  (\hat{q}^G+\mathbb{I})_{\rho\sigma}^{-1} m_{\sigma,G}^{(\nu)} -\frac{\alpha n \beta}{2}
-\beta (1-p)\sum_{\nu \rho}m_{\rho,B}^{(\nu)} \lambda_{\rho_B}^{(\nu)}\\
+p\ln \sum_{s_\rho} \exp\left\{\frac{\alpha\beta^2}{2}\sum_{\rho\neq\sigma}r_{\rho\sigma}^Gs^\rho s^\sigma\right\}
+(1-p)\mathbb{E}_B\left[\ln \sum_{s_\rho} \exp\left\{\frac{\alpha\beta^2}{2}\sum_{\rho\neq\sigma}r_{\rho\sigma}^Bs^\rho s^\sigma +\beta\sum_{\nu \rho} \lambda_{\rho_B}^{(\nu)}\xi_{B}^{(\nu)} s^\rho\right\}\right]
\end{multline}
\end{widetext}
Using Eq.~(\ref{eq:freenergy}), we can then write the free energy of this mixed model
\begin{equation}
f=-\lim_{n\rightarrow 0 } \frac{1}{n}A\left[q^B,q^G,r^B,r^G,m_B,m_G\right]\;,
\label{eq:free}
\end{equation}

At last we can recover the free energy in the replica symmetric case given by equation \ref{eq:freeE}. This is done by evaluating each term of (\ref{eq:free}) in this particular case and then by taking the limit of $n$ approaching zero just as prescribed by the replica method. 

\section*{Acknowledgement}
We thank Adriano Barra, Antonio Leonetti, Erik H\"ormann, Giancarlo Ruocco and Giorgio Parisi for the useful discussions.
NanoProbe,
LoTGlassy, Simons.

\bibliographystyle{unsrtnat}

\end{document}